\documentclass[11pt,a4paper]{article}
\usepackage{jheppub}

\usepackage{placeins}
\usepackage{amsmath,amssymb,mathtools,bm}
\usepackage{graphicx, color, hepunits}
\usepackage[dvipsnames,table]{xcolor}
\usepackage{cancel}
\usepackage[normalem]{ulem}
\usepackage{float}
\usepackage{filecontents}
\usepackage{multirow}
\usepackage{tikz}
\usepackage[caption=false]{subfig}
\usepackage{hyperref}
\usepackage{booktabs}
\usepackage{adjustbox}

\usetikzlibrary{decorations.pathmorphing}
\usetikzlibrary{arrows}
\usepackage{circuitikz}
\usetikzlibrary{arrows.meta}
\usetikzlibrary{shapes.misc}
\usetikzlibrary{positioning}
\usetikzlibrary{decorations.markings}

\usepackage[compat=1.1.0]{tikz-feynman}

\usepackage{hyperref} %Automatically links \label and \ref commands; Always load last
\hypersetup{
    colorlinks=true,       % false: boxed links; true: colored links
    linkcolor=blue,        % color of internal links
    citecolor=blue,        % color of links to bibliography
    filecolor=magenta,     % color of file links
    urlcolor=blue          % color of external links
}
\usepackage[utf8]{inputenc}
\usepackage[english]{babel}

\usetikzlibrary{shapes.misc}
\tikzset{cross/.style={cross out, draw=black, minimum size=2*(#1-\pgflinewidth), inner sep=0pt, outer sep=0pt},
cross/.default={1pt}}

\title{Two Higgs Doublet Solutions to the Strong CP Problem}

\author[a]{Quentin Bonnefoy,}
\author[b,c]{Lawrence Hall,}
\author[b,c]{Claudio Andrea Manzari}
\author[b,c]{and Bea Noether}

\affiliation[a]{Universit\'e de Strasbourg, CNRS, IPHC UMR7178, 23 rue du Loess, 67037 Strasbourg, France}

\affiliation[b,c]{Leinweber Institute for Theoretical Physics, University of California, Berkeley, CA 94720, U.S.A.}
\affiliation[b,c]{Theoretical Physics Group, Lawrence Berkeley National Laboratory, Berkeley, CA 94720, U.S.A.}

\emailAdd{qbonnefoy@unistra.fr}
\emailAdd{ljh@berkeley.edu}
\emailAdd{camanzari@berkeley.edu}
\emailAdd{bea\_noether@berkeley.edu}

\abstract{We solve the strong CP problem in a broad class of two Higgs doublet theories that will be probed at the Large Hadron Collider and at future colliders.\\
These theories feature CP and Abelian flavor symmetries, both broken softly in the scalar potential, that yield realistic quark masses and mixings.
The flavor symmetry charges are chosen so that $\bar{\theta}=0$ at tree level for all values of the Yukawa and quartic couplings of the theory. We prove that in all such theories the 1-loop contribution to $\bar{\theta}$ also vanishes, independently of the mass scale of the second Higgs doublet. We study two illustrative models with flavor group $\mathbb{Z}_3$. The direct contributions to the neutron electric dipole moment are negligible in both models.  While the 2-loop contribution to $\bar{\theta}$ is less than $10^{-12}$ in one model, it can be as large as $10^{-10}$ in the other, yielding the prospect of a signal in planned experiments.
Even with Abelian flavor symmetries, CP violation in neutral kaon mixing is generally expected to yield naturalness bounds on the masses of additional Higgs doublets of order 20 TeV. We prove that for all models in our class, where the flavor symmetry forces $\bar{\theta}$ to vanish at tree-level, the flavor-changing neutral currents are CP-conserving, yielding model-dependent bounds from neutral meson mixing near 1 TeV. 
Mixing of the two CP-even scalars gives corrections to the couplings of the 125 GeV Higgs state to $\bar{t}t, \bar{b}b, \bar{c}c, \bar{\tau}\tau$ and  $\bar{\mu} \mu$, giving possible signals at high luminosity runs at LHC and at future colliders. Furthermore, distinctive correlations between corrections in the various channels can probe the underlying flavor symmetry.

}

\begin{document}
\maketitle

% \tableofcontents
% \vspace{2cm}

\section{Introduction}

The strong CP problem, which refers to the absence of violation of the combination of Parity (P) and Charge-conjugation (C) in the strong interactions, is one of the most compelling puzzles in particle physics. The strength of the potential violation is described by the dimensionless parameter $\bar{\theta}$ which is constrained by the experimental limit on the neutron electric dipole moment (nEDM): $\bar{\theta} \leq 10^{-10}$~\cite{Pendlebury:2015lrz}. In the Standard Model (SM) of particle physics, the parameter $\bar{\theta}$ receives two distinct contributions. The first arises from the Quantum Chromodynamics (QCD) Lagrangian, which, constrained only by Lorentz and gauge invariance, includes a CP-odd term :  
\begin{equation}
\theta_{\rm QCD} \frac{g_s^2}{32\pi^2} G_a^{\mu\nu} \tilde{G}_{a,\mu\nu} \ ,
\end{equation}
where $G_a^{\mu\nu}$ is the gluon field strength tensor, $g_s$ is the strong coupling constant, and $\tilde{G}_{a,\mu\nu} = \frac{1}{2} \epsilon_{\mu\nu\alpha\beta} G_a^{\alpha\beta}$. The second contribution to $\bar{\theta}$ arises from the chiral transformations required to diagonalize the quark Yukawa matrices, $y^u$ and $y^d$, leading to  
\begin{equation}
\bar{\theta} = \theta_{\rm QCD} + \arg \det(y^u y^d) \ .
\end{equation}
These two contributions stem from entirely different SM sectors and have no intrinsic reason to cancel; the SM does not provide a theoretical understanding of the smallness of $\bar{\theta}$. 

Among the proposed strategies to solve this problem, one class of solutions invokes a fundamental CP~\cite{Georgi:1978xz, Mohapatra:1979kg, Nelson:1983zb, Barr:1984qx, Glashow:2001yz, Feruglio:2023uof, Feruglio:2024ytl, Hall:2024xbd, Feruglio:2025ajb} or P~\cite{Babu:1988mw,Babu:1989rb,Barr:1991qx, Hall:2018let, Dunsky:2019api, Bonnefoy:2023yoj} symmetry at high energies, which is either softly or spontaneously broken, having significant implications for the SM flavor sector. In such scenarios, CP or P ensures $\theta_{\rm QCD} = 0$, necessitating a mechanism that accounts for the large CP-violating phase observed in weak interactions while simultaneously explaining why $\arg \det(y^u y^d)$ remains small. While it could seem ad hoc to impose P or CP at the Lagrangian (or path integral) level in a QFT model, this finds an elegant realization in string theory, where they can arise as gauge symmetries and be spontaneously broken by the expectation values of moduli and matter fields~\cite{Strominger:1985it,Dine:1992ya,Choi:1992xp,Maiezza:2021dui}.

Motivated by early models addressing the strong CP problem \cite{Georgi:1978xz,Segre:1979dx,Mohapatra:1979kg, Glashow:2001yz}, some of us introduced a class of multi-Higgs doublet extensions of the SM that solves the strong CP problem~\cite{Hall:2024xbd}. (See Refs.~\cite{Nebot:2018nqn,Ferreira:2019aps,Ferro-Hernandez:2024snl} for other studies of multi-Higgs theories that consider the strong CP problem.) This framework is based on CP and a flavor symmetry, $G_{HF}$, both softly broken only in the scalar potential. 
%To achieve this, the scalar sector of the SM is extended to N copies of the Higgs doublet. 
Ref.~\cite{Hall:2024xbd} found many choices for $G_{HF}$ and for $G_{HF}$-charges of quarks and scalars that lead to fully realistic Yukawa coupling matrices describing the observed quark masses and mixing while giving $\bar{\theta} = 0$ at tree-level. In addition, it was found that radiative contributions to $\bar{\theta}$ vanished in all theories with two Higgs doublets, providing the mass of the second Higgs doublet is significantly larger than the electroweak scale. In this framework, the mass scale for the additional Higgs doublets ranges over many orders of magnitude and was typically taken far above the TeV scale, even at the cost of significant fine-tuning for the SM Higgs. The lower bound on this mass scale was not investigated.

In this paper, we prove that there is a large class of models with two Higgs doublets (2HDM), including the list presented in Ref.~\cite{Hall:2024xbd}, where there are no one-loop corrections to $\bar{\theta}$, regardless of the mass scale of the second Higgs doublet. Furthermore, we discuss radiative corrections to $\bar{\theta}$ at two-loops, presenting two illustrative models where these are well below the current experimental limit. Given that there is a theoretical advantage to having a low mass scale for the second Higgs doublet, namely that a low-scale UV completion of the theory allows much less fine-tuning than for a heavier Higgs, we study in this paper how light the second Higgs doublet can be. 

Since our theories do not satisfy the Glashow-Weinberg criterion \cite{Glashow:1976nt}, dimensional analysis suggests that the lower bound on the heavy Higgs mass from CP violation in neutral meson mixing is of order $10^5$ TeV. This turns out to be incorrect for two reasons.  First, a flavor symmetry restricts the relevant Yukawa couplings to be much less than unity, lowering the expected bound to around 20 TeV \cite{Hall:1993ca}. However, this is also a significant overestimate as we prove a surprising result: in all 2HDM with an abelian flavor symmetry that imposes $\bar{\theta}=0$ at tree-level, all neutral flavor-changing interactions are CP conserving. Hence, our results motivate experimental searches for theories with two Higgs doublets that solve the strong CP problem, both directly at current and future colliders and in flavor-changing processes as well as direct contributions to electric dipole moments. 

The paper is structured as follows. In Sec.~\ref{sec:SolTheory} we discuss the theories presented in Ref.~\cite{Hall:2024xbd}. In Sec.~\ref{sec:SPot} and Sec.~\ref{sec:YukCoup} we discuss the generalities of the scalar potential and Yukawa sector of theories with two Higgs doublets. The solution to the strong CP problem in these theories is studied in Sec.~\ref{sec:RadCorr}. In Sec.~\ref{sec:Models}  we present two illustrative models, showing that they account for the observed quark masses and mixings and have a two-loop contribution to $\bar{\theta}$ well below the experimental bound. In Sec.~\ref{sec:Pheno} we study experimental signals of the second Higgs doublet from direct contributions to the neutron electric dipole moment, neutral meson mixing and deviations from the SM in decays of the 125 GeV Higgs boson. We conclude in Sec.~\ref{sec:Conclusions}. In App.~\ref{app:MassMatrix}, we give the explicit form of the scalar mass matrix and in App.~\ref{app:Lag}, that of the scalar interactions with quarks. In App.~\ref{app:CPNC} we provide the proof of an important result of this paper, namely the absence of CP violation in the interactions of neutral scalars. Finally, in App.~\ref{app:2loops} we discuss two-loop radiative corrections to the nEDM.

\section{Multi-Higgs Solution to the Strong CP Problem}\label{sec:SolTheory}

Ref.~\cite{Hall:2024xbd} introduces a novel framework to solve the strong CP problem based on the following mild additions to the SM:
\begin{itemize}
    \item N copies of the scalar Higgs doublet,
    \item A CP symmetry and a flavor symmetry, $G_{HF}$, both softly broken in the scalar potential.
\end{itemize}
The phases in the scalar potential are transferred with opposite signs to the up- and down-quark Yukawa couplings. Additionally, the flavor symmetry dictates a texture for the Yukawa matrices with several zero entries, yielding realistic quark masses and CKM angles while enforcing $\arg \det(y^u y^d) = 0$. As a result, the strong CP problem is solved at tree level. Assuming a large separation between the mass scale $M$ of the additional scalars and the electroweak scale $v$, $M \gg v$, Ref.~\cite{Hall:2024xbd} demonstrates that one-loop corrections to $\bar{\theta}$ are proportional to the logarithm of the mass ratios of the heavy fields. This implies that in theories with two Higgs doublets (one light and one heavy) no one-loop corrections arise. Furthermore, it is shown that in theories with additional scalars, the corrections can remain below current experimental limits.

In the following, we focus on theories with two Higgs doublets and investigate whether this solution to the strong CP problem survives as the mass scale of the second Higgs doublet is lowered towards the weak scale. We show that the continuous flavor symmetry $G_{HF} = U(1)$ is too strong, as it predicts the absence of CP violation in weak interactions. However, there is a large number of 2HDM with a discrete flavor symmetry.  For example, Ref.~\cite{Hall:2024xbd} lists the 24 possible Yukawa textures for realistic $\mathbb{Z}_3$ theories, subject to certain simplifying conditions\footnote{Precisely, theories were selected when all CKM mixings could be retrieved through a first-order diagonalization procedure. Other models that require higher-order terms in the transition from the flavor to the mass basis have not been systematically explored.}, that have $\bar{\theta} = 0$ at tree-level. Relaxing the simplifying conditions allows many more such theories.

\section{The Scalar Potential}\label{sec:SPot}

The most general scalar potential, with softly broken CP, in a theory with two copies of the Higgs boson, $\Phi_1$ and $\Phi_2$, is
\begin{align}
\begin{split}
    % V(\Phi_1,\Phi_2) &= \mu_{11}\Phi_1^{\dagger}\Phi_1 + \mu_{22}\Phi_2^{\dagger}\Phi_2 + \mu_{12}(e^{-i\alpha}\Phi_1^{\dagger}\Phi_2+e^{i\alpha}\Phi_2^{\dagger}\Phi_1) + \lambda_1(\Phi_1^{\dagger}\Phi_1)^2\\
    % &+ \lambda_2(\Phi_2^{\dagger}\Phi_2)^2  + 2\lambda_3(\Phi_1^{\dagger}\Phi_1)(\Phi_2^{\dagger}\Phi_2) + 2\lambda_4(\Phi_1^{\dagger}\Phi_2)(\Phi_2^{\dagger}\Phi_1)\\
    % &+ \lambda_5\bigg[(\Phi_1^{\dagger}\Phi_2)^2 + (\Phi_2^{\dagger}\Phi_1)^2\bigg]+\lambda_6\bigg[\Phi_1^{\dagger}\Phi_1\Phi_1^{\dagger}\Phi_2 + \Phi_1\Phi_1^{\dagger}\Phi_1\Phi_2^{\dagger}\bigg]\\
    % & + \lambda_7\bigg[\Phi_2^{\dagger}\Phi_2\Phi_2^{\dagger}\Phi_1 + \Phi_2\Phi_2^{\dagger}\Phi_2\Phi_1^{\dagger}\bigg]\,,
    V(\Phi_1,\Phi_2) &= \mu_{11}\Phi_1^{\dagger}\Phi_1 + \mu_{22}\Phi_2^{\dagger}\Phi_2 + \mu_{12}\left(e^{i\alpha}\Phi_2^{\dagger}\Phi_1+\text{h.c.}\right)\\ 
    &+ \lambda_1(\Phi_1^{\dagger}\Phi_1)^2+ \lambda_2(\Phi_2^{\dagger}\Phi_2)^2  + 2\lambda_3(\Phi_1^{\dagger}\Phi_1)(\Phi_2^{\dagger}\Phi_2) + 2\lambda_4(\Phi_1^{\dagger}\Phi_2)(\Phi_2^{\dagger}\Phi_1)\\
    &+ \lambda_5\bigg[(\Phi_1^{\dagger}\Phi_2)^2 + +\text{h.c.}\bigg]+\lambda_6\bigg[\Phi_1^{\dagger}\Phi_1\Phi_1^{\dagger}\Phi_2+\text{h.c.}\bigg]+ \lambda_7\bigg[\Phi_2^{\dagger}\Phi_2\Phi_2^{\dagger}\Phi_1 +\text{h.c.}\bigg]\,,
\end{split}
\label{eq:scalarPot}
\end{align}
where all parameters are real. CP is softly broken by the unique possible phase $\alpha\neq 0$ in the quadratic terms. Following Ref.~\cite{Hall:2024xbd}, we introduce a flavor symmetry $G_{HF}$, under which $\Phi_1$ and $\Phi_2$ have different charges. In this work, we focus on abelian, continuous and discrete, flavor symmetries. Requiring the flavor symmetry to be only softly broken generally imposes $\lambda_5=\lambda_6=\lambda_7=0$. (Some non-abelian flavor symmetries and specific charge assignments under discrete abelian ones allow for $\lambda_5\neq 0$, see e.g. \cite{Segre:1979dx,Nebot:2018nqn}.) Therefore, in these theories, CP and $G_{HF}$ are only explicitly broken by the mixed mass term in the scalar potential% (the only operator in the lagrangian with dimension less than four)
. The vacuum expectation values (vevs) for $\Phi_1$ and $\Phi_2$ take the form~\cite{Ferreira:2004yd,Barroso:2005sm}
\begin{align}
    \langle \Phi_1 \rangle = \frac{1}{\sqrt{2}}\begin{pmatrix}
        0\\v_1
    \end{pmatrix}\,,\quad
    \langle \Phi_2 \rangle = \frac{1}{\sqrt{2}} \begin{pmatrix}
        0\\e^{i\theta} v_2
    \end{pmatrix}\,,
\end{align}
where we used the freedom of performing a hypercharge rotation to remove the phase in $\langle \Phi_1 \rangle$. 
Minimizing the scalar potential for negative $\mu_{12}$ and non-zero $v_{1,2}$, we find
\begin{align}
\begin{split}
    \theta &= \alpha\,,\\
    \mu_{11} &= -\frac{v_2}{v_1}\mu_{12} - v_1^2\lambda_1 - v_2^2(\lambda_3+\lambda_4)\,,
    \\
    \mu_{22} &= -\frac{v_1}{v_2}\mu_{12} - v_2^2\lambda_2 - v_1^2(\lambda_3+\lambda_4)\, .
\end{split}
\label{eq:VacuumMin}
\end{align}
%and $\mu_{12}$ is unconstrained.
Once $\mu_{12}$ is given, and given the value of the electroweak vev $v=\sqrt{v_1^2+v_2^2}=246$ GeV, the last two relations constrain the values of $\mu_{11}$ and $\mu_{22}$.

%\noindent\textbf{\large Higgs Basis}\\

It is convenient to move from the flavor basis $(\Phi_1, \Phi_2)$ to the so-called \textit{Higgs basis}  $(H_1, H_2)$, where the vev  $v%=\sqrt{v_1^2 + v_2^2}
$ is all in $H_1$, via
\begin{align}
\begin{pmatrix}
\Phi_1\\
\Phi_2
\end{pmatrix}
= U
\begin{pmatrix}
H_1\\H_2
\end{pmatrix}
\,,
\label{eq:Hbasis}
\end{align}
with 
\begin{align}
\label{eq:Umatrix}
    U = \begin{pmatrix}
\cos\beta&  \sin\beta \\
-\sin\beta \, e^{i\theta} & \cos\beta\, e^{i\theta}
\end{pmatrix}\,,
\end{align}
where $\cos{\beta} = v_1/v$ and $\sin{\beta} = -v_2/v$. We use the following component notation
\begin{align}
    H_1 = \begin{pmatrix}
G^+\\
\frac{v + H^0 + i\,G^0}{\sqrt{2}}
\end{pmatrix}\,,
\quad
H_2 = \begin{pmatrix}
H^+\\
\frac{R^0 + i\,I^0}{\sqrt{2}}
\end{pmatrix}\,,
\label{eq:Hbasis}
\end{align}
where $G^+$ and $G^0$ are the Goldstone bosons associated with electroweak symmetry breaking. In this basis, using the relations in Eq.~\eqref{eq:VacuumMin}, it can be shown that\footnote{Meanwhile, sanity checks can be performed : there is no term linear in $H_1$ nor $H_2$, so that the vevs in Eq.~\eqref{eq:Hbasis} are preserved, and there is no mixing between $G^+$ and $H^+$, nor between $G^0$ and $I^0$, since $G^+$ and $G^0$ exactly correspond to the Goldstone bosons associated to the breaking of $SU(2)_L\times U(1)_Y$.} the scalar potential becomes CP preserving, and therefore that there is no mixing between CP-odd ($I^0$) and CP-even eigenstates $(H^0, R^0)$. Therefore, the Higgs basis is the mass basis for $H^+$ and $I^0$, while there is mixing among the neutral states $H^0$ and $R^0$. We denote the CP-even mass eigenstates as $h$ and $H$, where $m_h$ is 125 GeV. The mass matrix is provided in App.~\ref{app:MassMatrix}.

\section{Yukawa Couplings and CP Conservation in Neutral Higgs Interactions} \label{sec:YukCoup}

The Yukawa couplings are
\begin{align} \label{eq:xxtilde}
    \mathcal{L}_Y = q^i\, x_{ij}^{\alpha}\, \bar{u}^j\, \Phi_{\alpha} + q^i\, \tilde{x}_{\alpha \, ik}\, \bar{d}^k\, \Phi^{* \, \alpha}\, + {\rm h.c.} \ ,
\end{align}
where we use two-component spinor notations. %All summations involve an upper and lower index, so that we can view  determined by the transformations of the corresponding quark and Higgs fields. In the entries that have non-zero charges, the actual coupling is zero, while entries with zero charge have non-zero couplings that are real by CP. 
Despite the fact that we impose an exact abelian flavor group $G_{HF}$ at the level of these Yukawa couplings, we treat $x_{ij}^{\alpha}$ and $\tilde{x}_{\alpha \, ik}$ as spurions and keep track of the flavor covariance of their (vanishing) spurious entries through our flavor space conventions, such that complete upper and lower index contractions yield a flavor singlet. The exact flavor symmetry implies that entries that have non-zero charges are actually zero, while entries with zero charge are non-vanishing but are real by CP. Moreover, since $\Phi_1$ and $\Phi_2$ have different flavor charges, each entry of the matrices $x^{\alpha} \Phi_\alpha$ and $\tilde{x}_{\alpha} \Phi^{* \,\alpha}$ has at most one contribution. 

Using the rotation in Eq.~\eqref{eq:Umatrix}, the Yukawa couplings in the Higgs basis read
\begin{align}
\label{eq:YukawaFlavorBasis}
    \mathcal{L}_Y = q^i\, x_{ij}^{\alpha}\, U_{\alpha \beta}\, \bar{u}^j\, H_{\beta} + q^i\, \tilde{x}_{\alpha \, ik}\,U^*_{\alpha \beta}\, \bar{d}^k\, H^*_{\beta}\, + {\rm h.c.} \ .
\end{align}
We do not use raised indices on $U$ since this matrix only exists when the flavor symmetry is broken (except for particular values of the scalar potential coefficients, each entry of $U$ receives contributions from several spurions of different flavor charges, mixed by the non-zero Higgs vevs). The Yukawa couplings to $H_1$ are
\begin{align} \label{eq:yud}
\begin{split}
    y^u_{ij} = x_{ij}^{\alpha}\, U_{\alpha 1}\ ,\quad y^d_{ik} = \tilde{x}_{\alpha \,ik}\, U^*_{\alpha 1}\,,
\end{split}
\end{align}
giving quark mass matrices $m_{u,d} = (v/ \sqrt{2})y^{u,d}$, and the Yukawa couplings to $H_2$ are 
\begin{align}\label{eq:zud}
\begin{split}
    z^u_{ij} = x_{ij}^{\alpha}\, U_{\alpha 2}\ ,\quad z^d_{ik} = \tilde{x}_{\alpha \,ik}\, U^*_{\alpha 2}\,.
\end{split}
\end{align}
%For later convenience we also define $z^u_{ij} = x_{ij}^{\alpha}\, U_{\alpha 2}$ and $z^d_{ik} = \tilde{x}_{\alpha \,ik}\, U^*_{\alpha 2}$, the Yukawa matrices for $H_2$. 
The $G_{HF}$ symmetry ensures that $\arg \det(y^u y^d) = 0$. We provide two illustrative models in Sec.~\ref{sec:Models}. For Model I the Yukawa matrices are
\begin{equation}\nonumber
    y^u =  
    \begin{pmatrix}
        x_{11}U_{11} & x_{12}U_{11} & x_{13}U_{21} \\
        x_{21}U_{21} & x_{22}U_{21} & 0 \\
        0 & 0 & x_{33}U_{11}
    \end{pmatrix}
    \; , \quad
    y^d =
    \begin{pmatrix}
        \tilde{x}_{11}U_{11}^* & 0 & 0 \\
        0 & \tilde{x}_{22}U_{21}^* & \tilde{x}_{23}U_{21}^* \\
        \tilde{x}_{31}U_{21}^* & \tilde{x}_{32}U_{11}^* & \tilde{x}_{33}U_{11}^*
    \end{pmatrix}
    \,,
\end{equation}
giving
\begin{align}
\begin{split}
&{\rm det}(y^u) = (x_{11}x_{22}x_{33}-x_{12}x_{21}x_{33})U_{11}U_{11}U_{21} \ ,\\
&{\rm det}(y^d) = (\tilde{x}_{11}\tilde{x}_{22}\tilde{x}_{33}-\tilde{x}_{11}\tilde{x}_{23}\tilde{x}_{32})U^*_{11}U^*_{11}U^*_{21}\ ,
\end{split}
\end{align}
so that indeed $\arg\det (y^uy^d) = 0$. 

For $G_{HF} = U(1)$, the phase in the scalar potential can be removed through a $G_{HF}$ transformation, since the soft scalar mass is the only operator in the theory that violates $G_{HF}$. Consequently, such models cannot account for the observed CKM phase and are not phenomenologically viable. 

On the other hand, if $G_{HF} = \mathbb{Z}_N$, the phase in the scalar potential is physical and cannot be eliminated by any field redefinition. In App.~\ref{app:CPNC} we prove that in all such theories where $\arg\det (y^uy^d)$ automatically vanishes at tree-level, a rephasing of the quark fields allows Yukawa interactions of all the neutral scalars to conserve CP. 
%In this case, we find that there exists a basis in which all neutral scalars have real couplings to quarks and the CP-violating phases reside in the CKM matrix and the couplings of the charged Higgs bosons. This result is crucial for the discussion that follows, and we have verified its validity by explicitly checking all $\mathbb{Z}_2$, $\mathbb{Z}_3$, and $\mathbb{Z}_4$ theories. \BN{Quentin's proof makes us able to say any $\mathbb{Z}_N$, no?}
Hence, there is a \textit{``phase basis''} where the Yukawa matrices, denoted with a bar, are all real and related as follows to the matrices in a generic Higgs basis :
\begin{align}
\begin{split}
    \bar{y}^{u,d} = P^{u,d}\, y^{u,d} \, P^{\bar{u},\bar{d}}\,,\\
    \bar{z}^{u,d} = P^{u,d}\, z^{u,d} \, P^{\bar{u},\bar{d}}\,,
\end{split}
\end{align}
where $P^{u,d,\bar{u},\bar{d}}$ are diagonal phase matrices. As indicated by the notation, left-handed up and down fields are rephased differently to reach the phase basis, which is thus no longer a Higgs basis. The above Yukawa matrices describe couplings of the neutral components of $H_{1,2}$ to quarks, while charged components couple via $P^d z^u P^{\bar u}$, etc. In this basis, all CP-violation originates from the CKM matrix and the couplings of the charged Higgs bosons. The presence of a CKM phase is permitted by $P^u \neq P^d$. The complete couplings of scalars and gauge bosons to quarks in this phase basis are shown in App.~\ref{app:Lag}. From the phase basis, the quark mass basis is reached by real rotations on the quark fields sending $\bar{y}^{u,d} \rightarrow \hat{y}^{u,d}$, which are real and diagonal, and $\bar{z}^{u,d} \rightarrow \hat{z}^{u,d}$, which are real and non-diagonal.

We stress the importance of the result of App.~\ref{app:CPNC}: in Sec. \ref{subsec:FCNC} we show that the absence of CP violation in tree-level neutral meson mixing reduces the bound on the heavy Higgs doublet from 20 TeV to around 1 TeV. Furthermore, in Sec. \ref{subsec:HiggsCouplings}, we show that the mixing of a CP-even state near a TeV with the 125 GeV Higgs leads to sizable deviations from SM couplings that could be observed at HL-LHC and future colliders. 

Ignoring some details, the proof of App.~\ref{app:CPNC} can be summarized as follows.
Comparing Eqs.~\eqref{eq:yud} and \eqref{eq:zud}, we see that we can obtain $z^{u,d}$ by taking the relation defining $y^{u,d}$ and replacing the first column of $U$ with the second column. Thus, from Eq.~\eqref{eq:Umatrix} we deduce that, while the magnitudes of $y^{u,d}_{ij}$ and $z^{u,d}_{ij}$ differ, they have the same phase. They also have the same pattern of zeros. Hence, all neutral Higgs couplings ($y^{u,d}$ and $z^{u,d}$) can be made real through quark phase rotations if and only if $y^{u,d}$ can be made real. Now, if such phase rotations on the quark fields %$u, \bar{u}, d, \bar{d}$ 
cannot be found, it must be that there is some rephasing-invariant product of Yukawa entries that has an imaginary part, either in the up or down sector (that are rotated independently). Without loss of generality, for a given $y$, the two relevant rephasing invariants %for either Yukawa matrix 
can be taken to be $y_{11} \, y_{22}\, y_{12}^*\,y_{21}^*\,$ and $y_{11}\, y_{22}\, y_{33}\, y_{12}^*\, y_{23}^*\, y_{31}^*\,$.
In App.~\ref{app:CPNC} we show that if either of these invariants has an imaginary part, then the requirement that the Yukawa matrix has non-zero determinant implies that more than one independent monomial of the $\Phi_\alpha$ contributes to the determinant, so that $\bar{\theta}$ is non-zero at tree-level, despite the restrictions due to $G_{HF}$. 

\section{Radiative Corrections to $\bar{\theta}$}\label{sec:RadCorr}

In the theories discussed above, $\bar{\theta}=0$ at tree-level. However, the experimental constraint from the nEDM is so strong that radiative corrections could be large enough to spoil the solution to the strong CP problem.
Here, we refer to radiative corrections to the quark mass matrices, that can induce $\bar{\theta}\neq 0$ through the quark field rotations needed to diagonalize them. The relation to keep in mind is 
\begin{equation}
\label{eq:FirstOrderTheta}
\bar{\theta} = {\rm arg\, det}\big[(m_u+\delta m_u)(m_d+\delta m_d)\big]\approx {\rm arg\, det}\big(m_um_d\big)+\text{Im Tr}\left(m_u^{-1}\delta m_u+m_d^{-1}\delta m_d\right) \ ,
\end{equation}
where $\delta m_{u(d)}$ encode the higher-order corrections and $\approx$ only holds at first order.

In Sec.~\ref{sec:SPot} and
Sec.~\ref{sec:YukCoup}, it was argued that in the phase basis, phases are only found in the CKM matrix and in the couplings of the charged Higgs boson. This result reduces tremendously the number of diagrams that have to be checked to estimate the quantum corrections to the relation $\bar{\theta} = 0$. Ellis and Gaillard~\cite{Ellis:1978hq} have already proved that, in the SM, threshold corrections to $\bar{\theta} = 0$ do not appear below three-loops and are of the order $\bar{\theta}\sim10^{-16}$. 

\subsection{One loop}

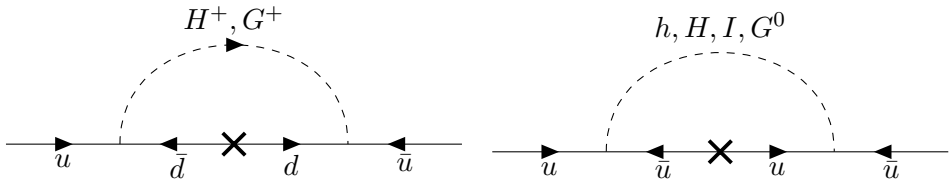
\begin{figure}
\centering    
\begin{tikzpicture}
    \begin{feynman}
        \vertex (L) at (-3,0);
        \vertex (R) at (+3,0);
        \vertex (M) at (0,0);
        \vertex (T) at (0,1.5);
        \vertex (LC) at (-1.5,0);
        \vertex (RC) at (+1.5,0);
        %\vertex (B) at (0,-2) {$\langle H_1 \rangle$};
        \node[cross out, draw, minimum size=7pt, inner sep=0pt, line width=0.5mm] at (0, 0) {};
       \diagram*{
            (L) -- [fermion,edge label' = {$u$}] (LC),
            (LC) -- [anti fermion, edge label' = {$\bar d$}] (M),
            (M) -- [fermion, edge label' = {$d$}] (RC),
            (RC) -- [anti fermion, edge label' = {$\bar u$}] (R),
            (LC) -- [charged scalar, half left, edge label={$H^+, G^+$}] (RC)
            };
    \end{feynman}
    \end{tikzpicture}\quad
       \begin{tikzpicture}
    \begin{feynman}
        \vertex (L) at (-3,0);
        \vertex (R) at (+3,0);
        \vertex (M) at (0,0);
        \vertex (T) at (0,1.5);
        \vertex (LC) at (-1.5,0);
        \vertex (RC) at (+1.5,0);
        %\vertex (B) at (0,-2) {$\langle H_1 \rangle$};
        \node[cross out, draw, minimum size=7pt, inner sep=0pt, line width=0.5mm] at (0, 0) {};
        \diagram*{
            (L) -- [fermion,edge label' = {$u$}] (LC),
            (LC) -- [anti fermion, edge label' = {$\bar u$}] (M),
            (M) -- [fermion, edge label' = {$u$}] (RC),
            (RC) -- [anti fermion, edge label' = {$\bar u$}] (R),
            (LC) -- [scalar, half left, edge label={$h, H, I, G^0$}] (RC)
            };
    \end{feynman}
    \end{tikzpicture}
\caption{One-loop radiative correction to the up-quark mass matrix due to the exchange of a scalar boson.}
\label{fig:oneloopFD}
\end{figure}

At one-loop, the only diagram that could induce a phase in ${\rm det}\big[(y^u+\delta y^u)(y^d+\delta y^d)\big]$ involves the exchange of a charged boson and is shown on the left panel of Fig.~\ref{fig:oneloopFD}.
The result reads\footnote{For more readability, we only included in Eq.~\eqref{eq:oneLoopYu} the first order in a perturbative expansion in terms of the quark masses, but all results that follow also hold when one deals with the complete result. See Sec. 3 of Ref.~\cite{Ferreira:2019aps} for an explicit example.}
\begin{align}
\label{eq:oneLoopYu}
\delta y^u = \sum_{\delta}\frac{\tilde{x}^{\alpha}\, \tilde{x}^{\dagger \beta}x^{\gamma}}{16\pi^2}\,  U^*_{\alpha \delta}\, U_{\beta 1}\, U_{\gamma \delta}\left[1+\frac{1}{\epsilon}+\log\left(\frac{\mu^2}{m_\delta^2}\right)\right]\ ,
\end{align}
where $m_\delta$ are the masses of the two charged bosons (the physical charged Higgs and the charged Goldstone boson). Using the unitarity of the U matrix, this result can be written as
\begin{align}
\begin{split}
\delta y^u &= \frac{\tilde{x}_{\alpha}\, \tilde{x}^{\dagger \beta}x^{\alpha}}{16\pi^2}\, U_{\beta 1}\, \left[1+\frac{1}{\epsilon}+\log\left(\frac{\mu^2}{m_2^2}\right)\right] + \frac{y^d\, y^{d\,\dagger}y^u}{16\pi^2}\,  \log\left(\frac{m_2^2}{m_1^2}\right)\,.
\end{split}
\label{eq:Hplusresult}
\end{align}
The term $y^d\, y^{d\,\dagger}y^u$ cannot induce $\bar{\theta}$ as it is an hermitian matrix times $y^u$ (recall Eq.~\eqref{eq:FirstOrderTheta}). The other term cannot affect the phase of the determinant of $y^u$ either. Indeed, each entry of the matrix  $(\tilde{x}_{\alpha}\, \tilde{x}^{\dagger \beta}x^{\alpha})_{ij}$ has the same free indices as $x^\beta_{ij}$ and hence the same flavor charges. Only the entries of these two matrices that have zero flavor charge can be non-zero, and in both cases the non-zero entries are real. 
%(and, being formed out of real matrices, it is real). This follows from the flavor symmetry : a zero in the $x,\tilde x$ matrices arises whenever the associated entry is charged under the flavor symmetry, while $\left(\tilde{x}^{\alpha}\, \tilde{x}^{\dagger,\, \beta}x^{\alpha}\right)_{ij}$ has the same flavor charges as the corresponding $x^\beta_{ij}$. Therefore, it is charged whenever $x^\beta_{ij}$ vanishes and must vanish as well, since the only non-zero objects forming $\tilde{x}^{\alpha}\, \tilde{x}^{\dagger,\, \beta}x^{\alpha}$ are flavor-neutral and can only combine to form more neutral objects. %\Qnote{I wonder what happens with this argument when some zeros are not 100\% due to flavor but also to hypercharge, as in the 3HDM Glashow model. Claudio tells me that this never happens for the $Z_3$ 2HDM theories}
Now, the pattern of zeros in the $x^\beta,\tilde x_\beta$ matrices, combined with the fact that they are real, was sufficient to obtain vanishing $\arg\det (y^uy^d)$ at tree-level after contracting the matrices with $U_{\beta 1}^{(*)}$. We thus see that this pattern survives when the first term of \eqref{eq:Hplusresult} is included. Hence, we conclude that, in the class of theories studied in this paper, there are no one-loop contributions to $\bar{\theta}$ due to the exchange of a charged boson, regardless of the Abelian flavor symmetry imposed as long as it ensures ${\rm{arg\, det}}\left(y^uy^d\right) =0$ at tree-level.\footnote{We remind the reader that we restrict our attention to scenarios in which the last equality arises solely from the structure of the Yukawa matrices, and not from the magnitude of their elements; that is, each contribution to the determinants carries the same phase. We do not consider finely-tuned cases in which the equality results from cancellations between different contributions to the determinants. An example of an alternative mechanism, where the cancellation arises due to the particular magnitudes of the Yukawa entries related by a discrete non-abelian symmetry, can be found in Ref.~\cite{Segre:1979dx}.} There is a contribution to $\bar{\theta}$ proportional to the product of the two terms in Eq.~\eqref{eq:Hplusresult}, corresponding to the second order neglected in the last equality in Eq.~\eqref{eq:FirstOrderTheta}, but we find it to be negligible.

In the large mass limit for the additional Higgs boson, corresponding to the large $m_2$ limit of the formulae above, we recover a result of Ref.~\cite{Hall:2024xbd}. We argue here that it holds independently of the mass scale of the second Higgs doublet. As the doublet becomes lighter, radiative corrections to the Yukawa couplings proportional to $\lambda v^2 / m_2^2$ become relevant and may, in principle, contribute to $\bar{\theta}$. Such corrections are included in the one-loop exchange of a neutral scalar, shown in the right panel of Fig.~\ref{fig:oneloopFD}, once the proper mass eigenstates run in the loop. In the unbroken phase, the corresponding diagrams involve insertions of quartic couplings along the scalar line, that match in the large $m_2$ limit onto higher-dimensional effective operators with three or more Higgses. However, as previously discussed, the scalar potential in the Higgs basis is CP-conserving, implying that all quartic couplings are real. Furthermore, in the phase basis, which shares the same scalar potential as the Higgs basis, the couplings of all neutral scalars are also real. Consequently, one-loop diagrams involving neutral scalar exchange cannot induce CP violation. %Additionally, the flavor structure ensures that these contributions have the same matrix structure as $y^{u(d)}$, and therefore cannot affect $\arg \det (y^u y^d)$.

%In practice, the induced $\delta y^u$ is structurally identical to that of Eq.~\eqref{eq:Hplusresult}, with only $x$ for up quarks (and $\tilde x$ for down quarks) in the first term and only one type of Yukawa matrix in the second term. Thus, as for the charged Higgs case, the flavor structure ensures that these contributions cannot affect $\arg \det (y^u y^d)$. The fact that there is no additional imaginary factor is due to the absence of mixing between CP-even and CP-odd Higgses.

\subsection{Two loops}\label{sec:twoLoopsTheta}

At two loops, there are many more diagrams that could contribute to $\bar{\theta}$. We work in the phase basis, where the quark Yukawa couplings are real but not diagonal, and the phases are in the couplings of the charged Higgs bosons and in the CKM matrix. The diagrams that could contribute to $\bar{\theta}$ involve two scalar loops and we show them in App~\ref{app:2loops}. The general form of the amplitude of each diagram is:
\begin{align} \label{eq:2loopthetagen}
    \mathcal{M} = \frac{\mathcal{A}}{(16\pi^2)^2}\bigg(1+\frac{1}{\epsilon}+f(m_1,m_2)\bigg)\,,
\end{align}
where $\mathcal{A}$ is the flavor part of these diagrams and involves the product of five Yukawa matrices and $f(m_1,m_2)$ is a dimensionless function of the two scalar masses in the loops. Using the unitarity of the U matrix it is easy to show that the finite and divergent pieces of each diagram do not change the phase of the determinant of $y^u$ and $y^d$. We explicitly show this for one diagram in App~\ref{app:2loops}. For the mass dependent-part of the amplitude one would need to compute the two-loop integrals. Instead, we follow a different approach. In the following section, we present two illustrative $\mathbb{Z}_3$ theories and we estimate each contribution to $\bar{\theta}$ assuming $f(m_i,m_j)\sim 1$, finding that they are all naturally below the experimental limit. While a dedicated scan of two-loop contributions in all $\mathbb{Z}_3$ theories is beyond the scope of this work, we anticipate that they will generically be small enough to pass experimental constraints.

\section{Two Illustrative Realistic Models}\label{sec:Models}

For the sake of clarity, we present two models that are phenomenologically viable and have sufficiently small two-loop corrections to $\bar{\theta}$ to unambiguously solve the strong CP problem. For both models we take $G_{HF} = \mathbb{Z}_3$, and for Model I the charge assignments are
\[
    \begin{array}{c|cccccccccccc}
        & \Phi_1 & \Phi_2 & q_1 & q_2 & q_3 & \bar{u}_1 & \bar{u}_2 & \bar{u}_3 & \bar{d}_1 & \bar{d}_2 & \bar{d}_3 \\ \hline
        \mathbb{Z}_3 & 0 & 2 & 0 & 1 & 2 & 0 & 0 & 1 & 0 & 1 & 1
    \end{array} \, .
    \]
Given these charges, the Yukawa couplings are constrained to take the form
\begin{equation} \label{eq:Z3model1}
    q 
    \begin{pmatrix}
        x_{11}\Phi_1 & x_{12}\Phi_1 & x_{13}\Phi_2 \\
        x_{21}\Phi_2 & x_{22}\Phi_2 & 0 \\
        0 & 0 & x_{33}\Phi_1
    \end{pmatrix}
    \bar{u}
    \quad , \quad
    q 
    \begin{pmatrix}
        \tilde{x}_{11}\Phi_1^* & 0 & 0 \\
        0 & \tilde{x}_{22}\Phi_2^* & \tilde{x}_{23}\Phi_2^* \\
        \tilde{x}_{31}\Phi_2^* & \tilde{x}_{32}\Phi_1^* & \tilde{x}_{33}\Phi_1^*
    \end{pmatrix}
    \bar{d}\ ,
\end{equation}
so that, at first order when non-diagonal entries are  small,
\begin{align}
\begin{split}
&y_u \approx |x_{11}U_{11}| \ , \quad y_c \approx |x_{22}U_{21}| \ ,\quad y_t \approx |x_{33}U_{11}| \ ,\\
&y_d \approx |\tilde{x}_{11}U^*_{11}| \ , \quad y_s \approx |\tilde{x}_{22}U^*_{21}| \ ,\quad y_b \approx |\tilde{x}_{33}U^*_{11}| \ ,\\
&|V_{us}|\approx \frac{|x_{12} U_{11}| }{y_c} \ , \quad |V_{cb}|\approx \frac{|\tilde{x}_{23}U^*_{21}|}{y_b} \ ,\quad |V_{ub}|\approx \frac{|x_{13}U_{21}|}{y_t} \ ,\\
&|J|\simeq \, \bigg|\frac{x_{12}}{x_{22}}\frac{x_{13}}{x_{33}}\frac{\tilde{x}_{23}}{\tilde{x}_{33}}\sin \beta\, \sin 3\theta \bigg|\ ,\\
\end{split}
\end{align}
where $J={\rm Im}\left(V_{ij}V_{kl}V_{il}^*V_{kj}^*\right)$  is the Jarlskog invariant. The strong CP problem is solved at tree level since
\begin{equation}
{\rm det}(y^u) = (x_{11}x_{22}-x_{12}x_{21})x_{33}U_{11}^2U_{21} \ , \quad {\rm det}(y^d) = \tilde{x}_{11}(\tilde{x}_{22}\tilde{x}_{33}-\tilde{x}_{23}\tilde{x}_{32})U^{*2}_{11}U^*_{21} \ ,
\end{equation}
so that clearly $\text{arg det}(y^u)=-\text{arg det}(y^d)$. There are no one-loop corrections as discussed above.
In order to have ${\rm det}(y^u) = y_u\,y_c\,y_t$ and ${\rm det}(y^d) = y_d\,y_s\,y_b$ we need $|x_{21}U_{21}|\lesssim y_u/|V_{us}| \approx 5\times 10^{-5}$ and $|\tilde{x}_{32}U_{11}|\lesssim y_s/|V_{cb}| \approx 0.01$. With these, and assuming the functions $f$ of (\ref{eq:2loopthetagen}) are $\mathcal{O}(1)$, we find that the largest contributions at two-loops to $\bar{\theta}$ are
\begin{equation} \label{eq:2loopthetaZ3}
    \bar{\theta} \simeq
        10^{-10} \, \tilde{x}_{31}\,c \,\sin 3\theta, \text{ with } c=3\sin{\beta} \, \text{ or} -2\cos{\beta}\ ,
\end{equation}
where the two values for $c$ correspond to the largest contributions with different analytical structures. From the hierarchical structure of the down quark mass matrix, we expect $\tilde{x}_{31} \ll \tilde{x}_{33} \sim y_b \sim 10^{-2}$.
% \BN{I will put in the Jarlskog invariant for this theory and a part in the text explaining that $\delta_{CKM}$ is nonzero and there is enough parameter freedom to match it without spoiling $\bar\theta < 10^{-10}$.} To accommodate the CP violation in the CKM matrix, we must be able to reproduce the Jarlskog invariant. Writing
% \begin{align}
%     J =& -\frac{i}{2 F F'}\det ([M_uM_u^\dagger,M_dM_d^\dagger])
%     \\
%     F \equiv& (m_u^2-m_c^2)(m_c^2-m_t^2)(m_t^2-m_u^2) 
%     \\
%     F'\equiv& (m_d^2-m_s^2)(m_s^2-m_b^2)(m_b^2-m_d^2)
% \end{align}
% we can calculate $J$ directly from $y^u$ and $y^d$. For this specific theory we find, using the approximate matching above,
% \begin{align}
%     |J|\simeq& \, \bigg|\frac{x_{12}}{x_{22}}\frac{x_{13}}{x_{33}}\frac{\tilde{x}_{23}}{\tilde{x}_{33}}x_{22}\sin 2\beta \bigg|
% \end{align}

In Model II the charges under $G_{HF} = \mathbb{Z}_3$ are
\[
    \begin{array}{c|cccccccccccc}
        & \Phi_1 & \Phi_2 & q_1 & q_2 & q_3 & \bar{u}_1 & \bar{u}_2 & \bar{u}_3 & \bar{d}_1 & \bar{d}_2 & \bar{d}_3 \\ \hline
        \mathbb{Z}_3 & 0 & 1 & 0 & 1 & 2 & 1 & 2 & 1 & 1 & 0 & 1
    \end{array} \ ,
    \]
giving Yukawa coupling matrices
\begin{equation}\label{eq:Z3model2}
    q 
    \begin{pmatrix}
        0 & x_{12}\Phi_2 & 0 \\
        x_{21}\Phi_2 & x_{22}\Phi_1 & x_{23}\Phi_2 \\
        x_{31}\Phi_1 & 0 & x_{33}\Phi_1
    \end{pmatrix}
    \bar{u}
    \quad , \quad
    q 
    \begin{pmatrix}
        \tilde{x}_{11}\Phi_2^* & \tilde{x}_{12}\Phi_1^* & \tilde{x}_{13}\Phi_2^* \\
        0 & \tilde{x}_{22}\Phi_2^* & 0 \\
        \tilde{x}_{31}\Phi_1^* & 0 & \tilde{x}_{33}\Phi_1^*
    \end{pmatrix}
    \bar{d}\ .
\end{equation}
It follows that
\begin{align}
\begin{split}
&y_u \approx \bigg|\frac{x_{12}x_{21}}{x_{22}}\frac{U_{21}^2}{U_{11}}\bigg|\ ,\quad y_c \approx |x_{22}U_{11}|\ ,\quad y_t \approx x_{33}U_{11}\ ,\\
&y_d \approx |\tilde{x}_{11}U^*_{21}|\ ,\quad y_s \approx |\tilde{x}_{22}U^*_{21}|\ ,\quad  y_b \approx |\tilde{x}_{33}U^*_{11}|\ ,\\
&|V_{us}|\approx \bigg|\frac{x_{12} U_{21} }{y_c} - \frac{\tilde{x}_{12} U_{21}^*}{y_s}\bigg|\ , \quad |V_{cb}|\approx \frac{|x_{23}U_{21}|}{y_t}\ ,\quad |V_{ub}|\approx \frac{|\tilde{x}_{13}U_{21}^*|}{y_b}\ ,\\
&|J|\simeq \, \bigg|\bigg(\frac{x_{12}\sin\beta}{x_{22}} - \frac{\tilde{x}_{12}}{\tilde{x}_{12}\sin\beta}\bigg)\frac{\tilde{x}_{13}}{\tilde{x}_{33}}\frac{x_{23}}{x_{33}}\sin^2 \beta\, \sin 3\theta \bigg|\ ,
\end{split}
\end{align}
and
\begin{equation}
{\rm det}(y^u) = x_{12}(x_{23}x_{31}-x_{21}x_{33})U_{11}U_{21}^2\ ,\quad {\rm det}(y^d) = \tilde{x}_{22}(\tilde{x}_{11}\tilde{x}_{33} -\tilde{x}_{13}\tilde{x}_{31})U^*_{11}U^{*2}_{21}\ .
\end{equation}
Using the same naturalness argument as before, we expect $|x_{21}U_{21}| \sim y_u/|V_{us}|,\; |x_{31}U_{11}| \sim y_u/|V_{us}V_{cb}|,\; |\tilde{x}_{31}U_{11}| \sim y_d/|V_{ub}|$ and we find that the two largest contributions to $\bar{\theta}$ at two-loops are
\begin{equation}
    \bar{\theta} \simeq 1\times 10^{-12}\, c  \ , \text{ with } c=1 \, \text{ or} \, \cot^2 \beta \ ,
    %\bigg(\frac{\cos\beta}{\sin\beta}\bigg)^2\,.
\end{equation}
%
%This prediction becomes even smaller if some of the parameters are taken to be below their naturally expected values.
so that the strong CP problem is solved.
Since $\cot \beta$ is unknown, $\bar{\theta}$ could be as large as the experimental bound of $10^{-10}$; in this model, planned nEDM experiments could discover a non-zero result. As we show below, such large values of $\cot \beta$ place a strong lower bound on the heavy Higgs doublet mass.

\section{Experimental Signals}\label{sec:Pheno}

Our analysis of the strong CP problem reveals theories that solve the strong CP problem for any value of the mass scale of the second Higgs doublet. It is therefore natural to ask how light the additional scalars can be and what are the most promising experimental signals for discovery.

\subsection{The Neutron Electric Dipole Moment}

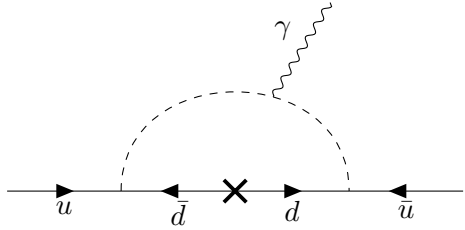
\begin{figure}
\centering  
\begin{tikzpicture}
    \begin{feynman}
        \vertex (L) at (-3,0);
        \vertex (R) at (+3,0);
        \vertex (M) at (0,0);
        \vertex (T) at (0,1.5);
        \vertex (LC) at (-1.5,0);
        \vertex (RC) at (+1.5,0);
        \vertex (g1) at (0.5,1.25);
        \vertex (g2) at (1.25,2.5);
        \node[cross out, draw, minimum size=7pt, inner sep=0pt, line width=0.5mm] at (0, 0) {};
        \diagram*{
            (L) -- [fermion,edge label' = {$u$}] (LC),
            (LC) -- [anti fermion, edge label' = {$\bar d$}] (M),
            (M) -- [fermion, edge label' = {$d$}] (RC),
            (RC) -- [anti fermion, edge label' = {$\bar u$}] (R),
            (LC) -- [ scalar, half left, edge label={}] (RC),
            (g1) -- [ boson, edge label={$\gamma$}] (g2)
            };
    \end{feynman}
\end{tikzpicture}
\caption{One-loop diagram that could induce an up-quark EDM.}
\label{nEDMscontributions}
\end{figure}

Given the stringent experimental bounds on the nEDM, and the prospect of significant improvements over the coming decade, it is crucial to verify whether our theories generate direct contributions, independent of $\bar{\theta}$, arising from quark EDMs and gluonic operators. Quark EDMs could be generated at one-loop from the diagrams in Fig.~\ref{nEDMscontributions} (recall that neutral Higgses do not mediate CP violation). We perform a perturbative diagonalization of the quark mass matrices, expanding up to second order in the off-diagonal elements, and evaluate the corresponding diagram. We find that no contributions to the up- or down-quark EDMs arise at this order in Models I and II. This result is further supported by analyzing the flavor invariants
\begin{equation}
    I_n = {\rm Tr}\bigg[z^d y^{d\,\dagger} z^u y^{u\,\dagger} (y^u y^{u\,\dagger})^n\bigg] \ .
\end{equation}
From Fig.~\ref{nEDMscontributions}, we see that they correspond to ${\rm Tr}\left[d_u y^{u\,\dagger} (y^u y^{u\,\dagger})^n\right]$, where $d_u$ is the electromagnetic dipole couplings of up quarks. Thus, in the quark mass basis\footnote{Strictly speaking, this is the result in the flavor basis in which $y^u$ is diagonal, which is aligned with the mass basis for up quarks, and in which all couplings of up quarks only are identical to mass basis quantities.} these invariants take the form $I_n \propto d_u m_u^{2n+1} + d_c m_c^{2n+1} + d_t m_t^{2n+1}$, with a real proportionality coefficient. Using the Yukawa structures discussed in Sec.~\ref{sec:Models}, one finds that each $I_n \in \mathbb{R}$, implying that $d_u$, $d_c$, and $d_t$ are all real. A similar result holds in the down-type quark sector. Quark EDMs could arise at the two-loop level but, in the two models of Sec.~\ref{sec:Models}, they do not induce a detectable nEDM. To show this, we recycled the approach followed in Sec.~\ref{sec:twoLoopsTheta} for the two-loop contributions to $\bar\theta$, since the flavor structure of the diagrams contributing to quark EDMs is identical (one can attach a photon line anywhere along the graphs of App.~\ref{app:2loops} to obtain the appropriate quark EDM diagrams). While we checked each contribution to $\bar\theta$, we also checked the contributions to quark EDMs.

Furthermore, as first observed by Weinberg~\cite{PhysRevLett.63.2333}, the CP-violating dimension-6 gluonic operator of the form
\begin{equation}
    \mathcal{O}_W = \frac{C_W}{\Lambda^2} G^{a\mu\nu} \tilde{G}^{b}_{\nu\rho} G^{c\rho}_{\mu} f^{abc}
\end{equation}
can also induce a sizable nEDM. However, in our class of theories, the additional neutral scalars do not mediate CP violation. As a result, the Weinberg operator is not generated below the three-loop level and can be safely neglected for all practical purposes.

Although direct contributions to the nEDM are negligible, in some models two-loop contributions to $\bar{\theta}$ may be large enough to give a signal in the coming decade in experiments that aim to improve the sensitivity by two orders of magnitude, as discussed in Sec. \ref{sec:Models}.

\subsection{Neutral Meson Mixing from Tree-Level FCNC} \label{subsec:FCNC}

Our solution to the strong CP problem predicts three additional scalars: a neutral CP-even, a neutral CP-odd and a charged one. $\Delta F = 2$ processes, such as $K$, $B_d$, $B_s$ and $D$ meson mixing are mediated at tree-level by the exchange of the neutral scalars. The most stringent experimental constraints are on the CP-violating coefficients of dimension-six $\Delta F = 2$ operators of the SM EFT. However, we have already shown that in our class of theories there is a basis where all neutral scalars have real couplings to quarks. Hence, our theories are much less constrained by FCNC than typical models with Abelian flavor symmetries \cite{Hall:1993ca}.  CP violation is mediated first at one-loop through the exchange of charged bosons. 

Using the operator analysis of Ref.~\cite{UTfit:2007eik}, the most general effective Hamiltonian for $\Delta S =2$ has the following form:
\begin{align}
  \label{eq:defheff}
  {\cal H}_{\rm eff}^{\Delta S=2}=\sum_{I=1}^{5} C_I\, Q_I +
  \sum_{i=I}^{3} \tilde{C}_I\, \tilde{Q}_I
\end{align}
with
\begin{equation}
\begin{aligned}
%Q_1 &= \bar d^{\alpha}_{jL} \gamma_\mu s^{\alpha}_{iL} \;\bar d^{\beta}_{jL} \gamma^\mu s^{\beta}_{iL} 
Q_1 &= d^{\alpha\dagger} \bar\sigma_\mu s^{\alpha} \; d^{\beta\dagger}\bar\sigma^\mu s^{\beta} \ ,
&
%Q_2 &= \bar d^{\alpha}_{jR} s^{\alpha}_{iL} \;\bar d^{\beta}_{jR} s^{\beta}_{iL} 
Q_2 &= s^{\alpha}\bar d^{\alpha}  \;s^{\beta} \bar d^{\beta}  \ ,
&
%Q_3 &= \bar d^{\alpha}_{jR} s^{\beta}_{iL} \;\bar d^{\beta}_{jR} s^{\alpha}_{iL} \;, \\
Q_3 &= s^{\beta}\bar d^{\alpha}  \;s^{\alpha}\bar d^{\beta} \ , \\
%Q_4 &= \bar d^{\alpha}_{jR} s^{\alpha}_{iL} \;\bar d^{\beta}_{jL} s^{\beta}_{iR} 
Q_4 &= s^{\alpha}\bar d^{\alpha}  \;\bar s^{\beta\dagger} d^{\beta\dagger}  \ , 
&
%Q_5 &= \bar d^{\alpha}_{jR} s^{\beta}_{iL} \;\bar d^{\beta}_{jL} s^{\alpha}_{iR} \;.
Q_5 &= s^{\beta}\bar d^{\alpha}  \;\bar s^{\alpha\dagger} d^{\beta\dagger} \ .
\end{aligned}
\end{equation}
In terms of our previous notation, $(u_i,d_i)=q_i$, while $\alpha$ and $\beta$ are colour indices. The operators
$\tilde{Q}_{1,2,3}$ are obtained from the
$Q_{1,2,3}%^{q_iq_j}
$ by the exchange $ L \leftrightarrow R$. An analogous formalism can be used for $\Delta B=2$ and $\Delta C =2$ processes. Without flavor symmetries, neutral-meson mixing places bounds on the masses of extra Higgs doublets far above the TeV scale. However, this is very significantly changed by flavor symmetries; for example, with approximate $U(1)^9$ flavor symmetries acting on $(q_i, \bar{u}_i, \bar{d}_i)$ \cite{Hall:1993ca}, current data gives bounds from $K, D$ and $B$ mixing that are all close to 1 TeV. However, in this framework the interactions of heavy neutral scalars violate CP and the bound from $\epsilon_K$ is about 20 TeV. In our theories there is no such tree-level contribution to $\epsilon_K$ opening up the possibility that $H$ and $I^0$ may be at the TeV scale. 

The coefficients $C_2, \tilde{C}_2$ and $C_4$ of Eq.~\eqref{eq:defheff} each receive contributions from all three neutral scalars. Moreover the two neutral CP-even scalars, $H^0$ and $R^0$, mix with angle $\alpha$, to become mass eigenstates $h$ and $H$. This induces flavor-changing Yukawa couplings of the SM-like Higgs boson, $h$. Hence, the couplings of mass eigenstate scalars to (up, down) quark mass eigenstates are
\begin{align}
    \mathcal{L} \supset 
    - \frac{1}{\sqrt{2}}(u,d) \left[ \left( c_\alpha \, \hat{y}^{u,d} + s_\alpha\, \hat{z}^{u,d} \right)h + \left( -s_\alpha\, \hat{y}^{u,d} + c_\alpha\, \hat{z}^{u,d} \right) H + i\, \hat{z}^{u,d}\, I^0 \right](\bar{u}, \bar{d})\, 
    + \text{h.c.},
\end{align}
where $c_\alpha = \cos \alpha$ and $s_\alpha = \sin \alpha$ and we remind the reader that $\hat{y}^{u,d}$ are diagonal matrices and $\hat{z}^{u,d}$ are the real Yukawa couplings of $H_2$ in the quark mass basis. At tree-level, the exchange of $h$, $H$ and $I^0$ generates 
%the Wilson coefficients $C_I$ and $\tilde{C}_I$ of (\ref{eq:defheff}) 
%Taking the heavy scalars to be aligned with $H_2$ (this is exact for the CP-odd scalar, while we neglect the mixing between the CP-even states), at tree-level we find
% \begin{align}
% \begin{split}
%     C_2(M) = -\frac{(\hat{z}^{u,d}_{ij})^2}{4 M^2},\qquad
%     C_4(M) = -\frac{\hat{z}^{u,d}_{ij}\hat{z}^{u,d}_{ji}}{2M^2},\quad
%     \tilde{C}_2(M) = -\frac{(\hat{z}^{u,d}_{ji})^2}{4 M^2}\,.
% \end{split}
% \end{align}

\begin{align}
\begin{split}
    C_2 &= -\frac{(\hat{z}^{u,d}_{ij}s_{\alpha})^2}{4 m_h^2} -\frac{(\hat{z}^{u,d}_{ij}c_{\alpha})^2}{4 m_H^2} + \frac{(\hat{z}^{u,d}_{ij})^2}{4 m_I^2} \simeq -\frac{(\hat{z}^{u,d}_{ij}s_{\alpha})^2}{4}\bigg(\frac{1}{m_h^2}-\frac{1}{m_H^2}\bigg)\,,\\
    C_4 &= -\frac{\hat{z}^{u,d}_{ij}\hat{z}^{u,d}_{ji}s_{\alpha}^2}{2 m_h^2} -\frac{\hat{z}^{u,d}_{ij}\hat{z}^{u,d}_{ji}c_{\alpha}^2}{2 m_H^2} -\frac{\hat{z}^{u,d}_{ij}\hat{z}^{u,d}_{ji}}{2 m_I^2} \simeq -\frac{\hat{z}^{u,d}_{ij}\hat{z}^{u,d}_{ji}}{2}\bigg(\frac{s_{\alpha}^2}{m_h^2} + \frac{2-s_{\alpha}^2}{m_H^2}\bigg)\,,
    \\
    \tilde{C}_2 &= -\frac{(\hat{z}^{u,d}_{ji}s_{\alpha})^2}{4 m_h^2} -\frac{(\hat{z}^{u,d}_{ji}c_{\alpha})^2}{4 m_H^2} + \frac{(\hat{z}^{u,d}_{ji})^2}{4 m_I^2} \simeq -\frac{(\hat{z}^{u,d}_{ji}s_{\alpha})^2}{4}\bigg(\frac{1}{ m_h^2}-\frac{1}{ m_H^2}\bigg)\,,
\end{split}
\label{eq:WCmesonmixing}
\end{align}
where $(i,j) = (12), (23),(13)$ for $K/D, B_s, B_d$ mixing. Since $H$ and $I^0$ are somewhat heavier than the weak scale, they are approximately degenerate with mass $M = m_{H} \simeq m_I$. 
% \sout{As discussed in the next subsection, we take the quartic scalar couplings small enough that corrections from the mixing of the CP-even scalars are sub-dominant.} 
The coefficients $C$ and $\tilde{C}$ are scaled down to the relevant energy scale and compared with experimental limits. For the numerical analysis, we include the matching corrections of Ref.~\cite{Buras:2012fs}, we use the 2-loop renormalization group evolution of Refs.~\cite{Ciuchini:1997bw, Buras:2000if}, the matrix element parametrizations given in Refs.~\cite{FlavourLatticeAveragingGroup:2019iem,FermilabLattice:2016ipl,Carrasco:2015pra} and the results of Ref.~\cite{UTfit:2007eik}. Since we expect $M$ to be at least a few times larger than $m_h$, it is clear from Eq.~\eqref{eq:WCmesonmixing} that constraints on $C_2$ and $\tilde{C}_2$ will affect $s_{\alpha}$, while constraints on $C_4$ will correlate the allowed magnitude of $s_{\alpha}$ and $M$. Assuming that we can use single-operator bounds\footnote{This is accurate if the bound from one operator dominates and approximates the result of a complete fit. We find that this holds in one of our models, as we show below. For the other one, it does not necessarily hold, and the contribution of $C_2,\tilde C_2,C_4$ to meson mixing should be consistently combined. Nevertheless, for simplicity, we simply impose the three bounds of Eq.~\eqref{eq:WCmesonmixing} on the model as a proxy for the exact result. When the strength of two bounds is comparable, destructive interference would make our result too conservative, while constructive interference would make our bound on $M$ a factor $\lesssim\sqrt 2$ too loose.}, the results of Ref.~\cite{UTfit:2007eik} for $K, D, B_d$ and $B_s$ mixing respectively give 
\begin{align}
\begin{split}
    % C_2 &\simeq -\frac{(\hat{z}^{u,d}_{ij}s_{\alpha})^2}{4 m_h^2} \lesssim \Biggl\{ \frac{1}{7.3\times 10^3\, {\rm TeV}},\, \frac{1}{2.5\times 10^3\, {\rm TeV}},\, \frac{1}{1.2\times 10^3\, {\rm TeV}},\, \frac{1}{130\, {\rm TeV}} \Biggr\}  \,,\\
    % C_4 &%\simeq -\frac{\hat{z}^{u,d}_{ij}\hat{z}^{u,d}_{ji}}{2}\bigg(\frac{s_{\alpha}^2}{m_h^2} + \frac{2-s_{\alpha}^2}{m_H^2}\bigg) 
    % \lesssim \Biggl\{ \frac{1}{17\times 10^3\, {\rm TeV}},\, \frac{1}{4.6\times 10^3\, {\rm TeV}},\, \frac{1}{2.2\times 10^3\, {\rm TeV}},\, \frac{1}{250\, {\rm TeV}} \Biggr\} \,,
    % \\
    % \tilde{C}_2 &\simeq -\frac{(\hat{z}^{u,d}_{ji}s_{\alpha})^2}{4 m_h^2} \lesssim \Biggl\{ \frac{1}{7.3\times 10^3\, {\rm TeV}},\, \frac{1}{2.5\times 10^3\, {\rm TeV}},\, \frac{1}{1.2\times 10^3\, {\rm TeV}},\, \frac{1}{130\, {\rm TeV}} \Biggr\}\,.
    \sqrt{|C_2|} &\simeq \frac{|\hat{z}^{u,d}_{ij}s_{\alpha}|}{2 m_h} \lesssim \Biggl\{ \frac{1}{7.3\times 10^3\, {\rm TeV}},\, \frac{1}{2.5\times 10^3\, {\rm TeV}},\, \frac{1}{1.2\times 10^3\, {\rm TeV}},\, \frac{1}{130\, {\rm TeV}} \Biggr\}  \,,\\
    \sqrt{|C_4|} &%\simeq -\frac{\hat{z}^{u,d}_{ij}\hat{z}^{u,d}_{ji}}{2}\bigg(\frac{s_{\alpha}^2}{m_h^2} + \frac{2-s_{\alpha}^2}{m_H^2}\bigg) 
    \lesssim \Biggl\{ \frac{1}{17\times 10^3\, {\rm TeV}},\, \frac{1}{4.6\times 10^3\, {\rm TeV}},\, \frac{1}{2.2\times 10^3\, {\rm TeV}},\, \frac{1}{250\, {\rm TeV}} \Biggr\} \,,
    \\
    \sqrt{|\tilde{C}_2|} &\simeq \frac{|\hat{z}^{u,d}_{ji}s_{\alpha}|}{2m_h} \lesssim \Biggl\{ \frac{1}{7.3\times 10^3\, {\rm TeV}},\, \frac{1}{2.5\times 10^3\, {\rm TeV}},\, \frac{1}{1.2\times 10^3\, {\rm TeV}},\, \frac{1}{130\, {\rm TeV}} \Biggr\}\,.
\end{split}
\label{eq:WCmesonmixing}
\end{align}
The couplings $\hat{z}^{u,d}_{ji}$ are model dependent, and hence so are the bounds on $s_{\alpha}$ and $M$. For Model I of Eq.~\eqref{eq:Z3model1} and Model II of Eq.~\eqref{eq:Z3model2}, the most stringent $\Delta F = 2$ bounds  result from $D$ mixing. We find
\[
\begin{array}{lclcl}
\text{Model I :}   &\quad& \hat{z}^u_{12} \simeq (\cot\beta + \tan\beta) V_{us} y_c \ , 
                  &\qquad& \hat{z}^u_{21} \simeq (\cot\beta + \tan\beta) V_{us} y_u\ , \\[1ex]
\text{Model II :}  &      & \hat{z}^u_{12} \simeq \dfrac{x_{12}}{\cos\beta} \approx (\cot\beta + \tan\beta)\sqrt{y_u y_c}\ , 
                  &      & \hat{z}^u_{21} \simeq \dfrac{x_{21}}{\cos\beta} \approx (\cot\beta + \tan\beta)\sqrt{y_u y_c}\ ,
\end{array}
\]
where for Model II we made the reasonable assumption $x_{12}\sin \beta\approx x_{21}\sin \beta \approx \sqrt{y_u y_c}$, since only their product is constrained by quark masses and mixing angles. We thus see that $\tilde C_2$ contributes negligibly to meson mixing in Model I, whereas it does as much as $C_2$ in Model II. The constraints on $C_2$ can be translated into
\begin{align}
\begin{split}
  {\rm Model\; I :}&\quad |s_{\alpha}(\tan\beta + \cot\beta)| \lesssim \sqrt{\frac{6\times 10^{-3} \, (M/{\rm TeV})^2}{(M/{\rm TeV})^2 - 2\times 10^{-2}} } \simeq 0.1 \ ,\\
  {\rm Model\; II :}&\quad \left|s_{\alpha}(\cot\beta + \tan\beta)\right| \lesssim \sqrt{\frac{(M/{\rm TeV})^2}{6(M/{\rm TeV})^2 - 0.1} } \simeq 0.4 \ ,
\end{split}
\label{eq:C2constraints}
\end{align}
while the constraints on $C_4$ yield
\begin{align}
\begin{split}
  {\rm Model\; I :}&\quad (\tan\beta + \cot\beta)^2\, \frac{(M/{\rm TeV})^2 s_{\alpha }^2+0.03}{(M/{\rm TeV})^2}\lesssim 0.4 \ ,\\
  {\rm Model\; II :}&\quad (\cot\beta + \tan\beta)^2\, \frac{64\,(M/{\rm TeV})^2 s_{\alpha}^2 + 2}{(M/{\rm TeV})^2}\lesssim 1.1 \ .
\end{split}
\label{eq:C4constraints}
\end{align}
Those $C_4$ bounds constrain in particular $|s_{\alpha}(\tan\beta + \cot\beta)|$ to be smaller than $\sqrt{0.4}\approx 0.6$ for Model I and $\sqrt{1.1/64}\approx 0.1$ for Model II. Thus, for the latter, the $C_4$ bound is sufficient.

\begin{figure}
    \centering
    \includegraphics[width=0.85\linewidth]{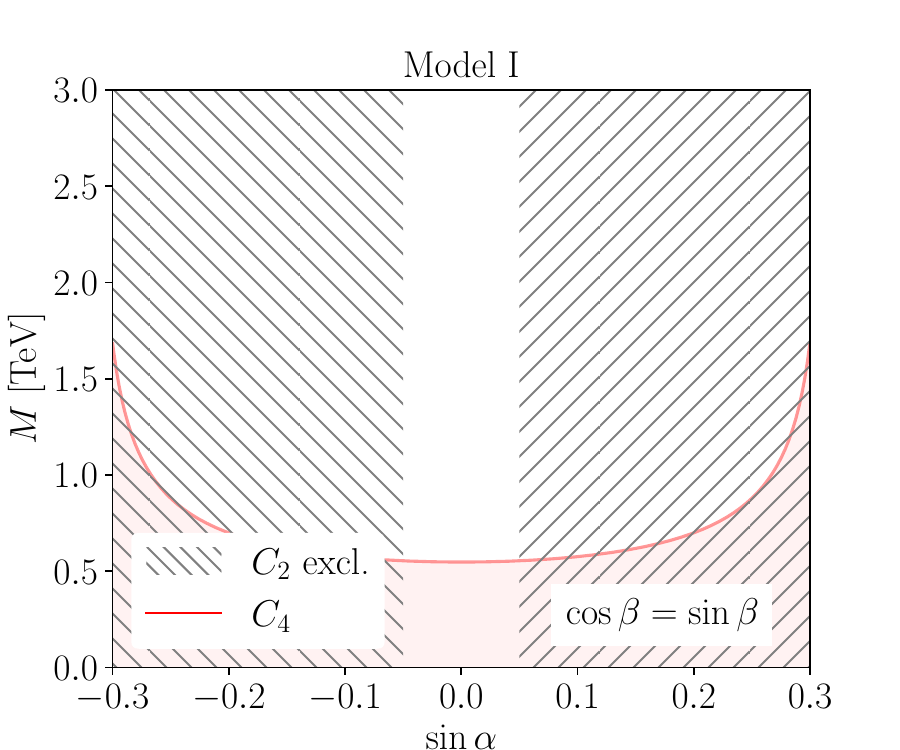}\\
    \includegraphics[width=0.85\linewidth]{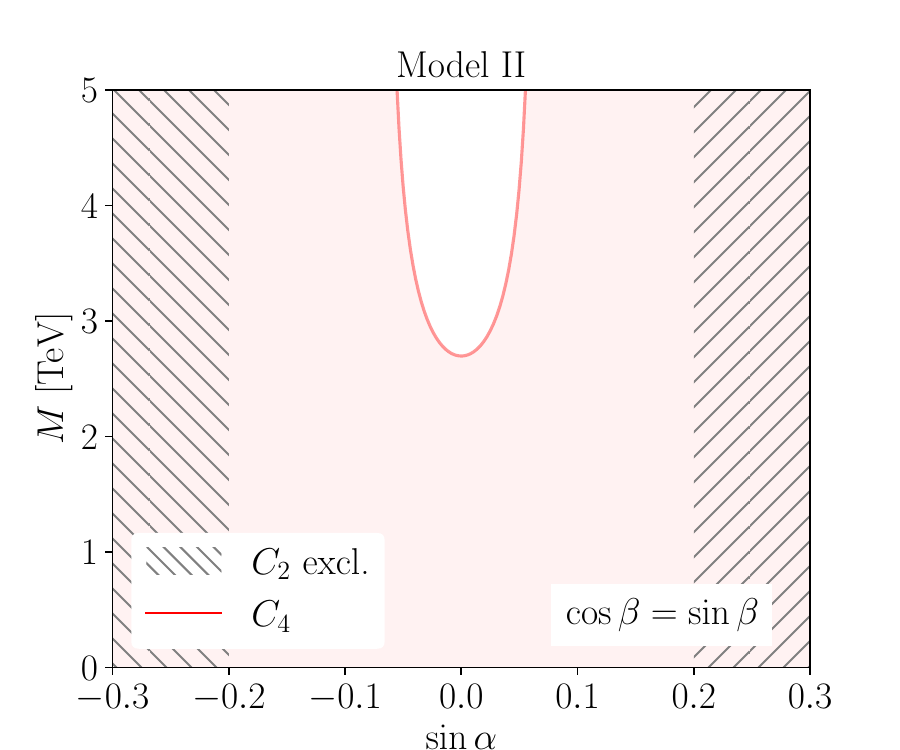}
    \caption{Allowed parameter space for $\sin\alpha$ and $M$ in the two models explicitly studied in this work, assuming $\cos\beta = \sin\beta$.%\BN{ For Model I we make the assumption that no cancellations between $C_2$ and $C_4$ happen.} 
    %For Model II we made the reasonable assumption $x_{12}\sin \beta\approx x_{21}\sin \beta \approx \sqrt{y_u y_c}$, since only their product is constrained by quark masses and mixing angles.
    }
    \label{fig:mesonmixing}
\end{figure}

The allowed parameter space for $s_\alpha$ and $M$, under the assumption $\cos\beta = \sin\beta$, is shown in Fig.~\ref{fig:mesonmixing}. In Model~I, the constraint on $s_\alpha$ from $C_2$ is often stronger than that from $C_4$, as a result of $\hat{z}_{12} \gg \hat{z}_{21}$. In contrast, for Model~II, where we assume $\hat{z}_{12} = \hat{z}_{21}$, the $C_4$ constraint is dominant due to tighter experimental limits, as noted above. Consequently, the allowed range for $M$ in Model~II is narrower than in Model~I, since the larger product of couplings entering $C_4$, $\hat{z}_{12}\hat{z}_{21}$, leads to stronger bounds. This highlights the model dependence of these constraints.
The bounds from $K$, $B_d$ and $B_s$ mixing in these two models are weaker.
% The bounds on $M$ from $K$ mixing in these two models are weaker,
% \begin{align} \label{eq:Kmixing}
%     M \; > \; \left(\frac{\tilde{x}_{31} /\tilde{x}_{32}}{x_{13}/x_{33}}( \tan \beta + \cot \beta ),\; 0.5\sqrt{1+ \tan^2 \beta} \right)  \; \mbox{TeV} \; \geq \;  (\approx 2, \;0.7) \; \mbox{TeV}
% \end{align}
% respectively.
% In Model I this bound depends on $\tilde{x}_{31}$ and $\tilde{x}_{32}$, which are not determined from quark masses and mixings. Taking natural values $\tilde{x}_{31} / \tilde{x}_{33} = x_{13}/x_{33}$ and $\tilde{x}_{32}=\tilde{x}_{23}$, and taking $\tan \beta =1$, gives the final numerical results. The bounds from $B$ mixing are weaker, about 300 GeV.
Although these results apply only to Models I and II, it appears likely that, in our general class of 2HDM, the lower bound on $M$ from FCNC constraints will be close to a TeV, as for theories with approximate $U(1)^9$ flavor symmetry. 

%However, some models can have strong departures from this typical behavior.  For example, in a $Z_3$ model with Yukawa coupling textures
%
%\begin{equation}\nonumber
%    \begin{pmatrix}
%        \Phi_2 & \Phi_2 & 0 \\
%        \Phi_2 & \Phi_2 & 0 \\
%        0 & 0 & \Phi_1
%    \end{pmatrix}
%    \bar{u}
%    \quad , \quad
%    q 
%    \begin{pmatrix}
%        \Phi_1^* & 0 & \Phi_2^* \\
%        \Phi_1^* & 0 & \Phi_2^* \\
%        0 & \Phi_2^* & \Phi_1^*
%    \end{pmatrix}
%    \bar{d}\,,
%\end{equation}
%the tree-level contribution to D meson mixing is absent as the matrix $\hat{z}^u$ is diagonal.

\subsection{Yukawa Couplings of the 125 GeV Higgs Boson}
\label{subsec:HiggsCouplings}

% The two neutral CP-even scalars mix, with angle $\alpha$, inducing a misalignment between the quark Yukawa couplings and the couplings of the SM-like Higgs boson to fermions. Hence, the couplings of CP-even mass eigenstate scalars to (up, down) quark mass eigenstates are
% %
% \begin{align}
%     \mathcal{L} \supset 
%     - \frac{1}{\sqrt{2}}(u,d) \left[ \left( c_\alpha \, \hat{y}^{u,d} + s_\alpha\, \hat{z}^{u,d} \right)h + \left( -s_\alpha\, \hat{y}^{u,d} + c_\alpha\, \hat{z}^{u,d} \right) H\right](\bar{u}, \bar{d})\, 
%     + \text{h.c.},
% \end{align}
% %
% where $c_\alpha = \cos \alpha$ and $s_\alpha = \sin \alpha$. 
Since the second CP-even scalar has a mass $M$ typically at or above the TeV scale, the mixing angle $\alpha$ can be treated perturbatively. Using (\ref{eq:CPevenmassmatrix}) we find
\begin{align}
    \sin\alpha \; \simeq \; \cos\beta\sin\beta \, \Biggl(\cos^2\beta \, (\lambda_1-\lambda_3-\lambda_4) + \sin^2\beta \, (-\lambda_2+\lambda_3+\lambda_4)\Biggr)\, \frac{v^2}{M^2} \ .
    \label{eq:mix_pert}
\end{align}
%
%App.~\ref{app:MassMatrix}. 
In typical models, contributions to neutral meson mixing amplitudes have some terms proportional to $\tan \beta$ and others to $\cot \beta$ and hence the mildest limit on $M$ occurs in the region where $\tan \beta \simeq 1$. 
A key question is the size of the various quartic couplings, $\lambda_a$, and the combinations $(\lambda_1-\lambda_3-\lambda_4)$ and $(-\lambda_2+\lambda_3+\lambda_4)$ in particular. In the two models studied above, the constraints on $\sin\alpha$ from meson mixing give an approximate upper bound of $\lambda \lesssim M^2/{\rm TeV}^2$. Other theories are likely to lead to similar results, so that the quartic couplings can be $\mathcal{O}(1)$ for heavy scalars at the TeV scale. As we discuss in the following, with the results in Eqs.~\eqref{eq:C2constraints}-\eqref{eq:C4constraints}, deviations from SM behavior of the 125 GeV state are not expected to be seen at LHC but can be discovered at future colliders.

The effective flavor-diagonal Higgs-fermion couplings are conventionally parametrized as
\begin{align}
    \mathcal{L} \supset -\frac{m^f_i}{v} \, \bar{f}_i \left( \kappa^f_i + i\, \tilde{\kappa}^f_i \gamma_5 \right) f_i\, h\,,
\end{align}
where $f=u,d,e$ for up, down and charged lepton sectors and  $(\kappa^f_i, \tilde{\kappa}^f_i) = (1, 0)$ in the SM. %In our framework  
For the class of theories discussed in this paper, $\tilde{\kappa}^f_i = 0$ and deviations from the SM predictions take the form
\begin{align}
    \kappa^{u,d,e}_i \simeq \cos\alpha + \sin\alpha \;\frac{\hat{z}^{u,d,e}_{ii}}{\hat{y}^{u,d,e}_{ii}}\,,
\end{align}
where we include charged-lepton interactions.
If the couplings $(x_{ii}, \tilde{x}_{ii})$ of Eq.~\eqref{eq:xxtilde}, and similar couplings for charged leptons, arise from an interaction with $\Phi_1$, then $\hat{z}^{u,d.e}_{ii} /\hat{y}^{u,d,e}_{ii} = \tan \beta$; if they arise from $\Phi_2$, then $\hat{z}^{u,d,e}_{ii} /\hat{y}^{u,d,e}_{ii} = -\cot \beta$. For Model I of Sec.~\ref{sec:Models}, one finds $\tan\beta$ for up, top, down and bottom quarks, and $-\cot\beta$ for charm and strange quarks. For the first and second generations some models, like Model II of Sec.~\ref{sec:Models}, will have some $(x_{ii}, \tilde{x}_{ii}) =0$. In this case the corresponding diagonal entries of $\hat{z}$ and $\hat{y}$ arise from rotations to diagonalize $y^{u,d,e}$, so that $\hat{z}^{u,d,e}_{ii} /\hat{y}^{u,d,e}_{ii}$ could be $\tan \beta, \,\cot \beta, \, (2\tan \beta + \cot\beta)$ or $(-2\cot\beta - \tan\beta)$. (If more than one rotation contributes to make $x_{ii},\tilde x_{ii}$ non zero, more values for the ratio can arise, but this is unlikely in theories with flavor hierarchies.) Hence, the magnitudes of deviations from SM couplings take one of four possible forms
\begin{align}
    |\kappa^{u,d,e}_i - 1| \; \simeq \; \Big|\cos \alpha - 1 + \sin \alpha \;\Big(\tan \beta, \;\cot \beta, \; 2\tan \beta + \cot\beta, \; -2\cot\beta - \tan\beta \Big)\Big| \ . 
    %\qquad n,m = 0,\pm 1, \pm 2.
    \label{eq:kappa_minus_1}
\end{align} 
We are unable to predict the sign of $(\kappa^{u,d,e}_i-1)$.
If more than one deviation can be measured, both $\tan \alpha$ and $\tan \beta$ can be extracted from data. The characteristic feature of our theories is that the ratio of any two deviations is just a low power of $\tan \beta$. For instance, in Model II, we have (when $\alpha$ is small) $|\kappa^{u}_{2,3} - 1| \simeq |\kappa^{d}_{3} - 1| \simeq |\sin \alpha \,\tan \beta|$, $|\kappa^{d}_{1,2} - 1| \simeq |\sin \alpha \,\cot \beta|$ and $|\kappa^{u}_1 - 1| \simeq |\sin \alpha (2\cot\beta+\tan \beta)|$.
%, where we have taken $\sin \alpha$ small.
%
\begin{table}[t]
\begin{adjustbox}{max width=\textwidth}
\renewcommand{\arraystretch}{1.5}
\begin{tabular}{l|c|c|c|cc|ccc|ccc|c|cc}
\toprule
 & LHC & HL-LHC & LHeC & \multicolumn{2}{c|}{HE-LHC} & \multicolumn{3}{c|}{ILC} & \multicolumn{3}{c|}{CLIC} & CEPC & \multicolumn{2}{c}{FCC-ee} \\
& ATLAS~\cite{ATLAS:2022vkf}& & & S2 & S2' & 250 & 500 & 1000 & 380 & 1500 & 3000 & & 240 & 365 \\
 &  & \% & \% & \% & \% & \% & \% & \% & \% & \% & \% & \% & \% & \% \\
\midrule
\rowcolor[gray]{0.9}
$\kappa_{c}$ & $0.03^{+3.02}_{-0.03}$ & -- & 4.1 & -- & -- & 2.5 & 1.3 & 0.9 & 4.3 & 1.8 & 1.4 & 2.2 & 1.8 & 1.3 \\
$\kappa_{t}$ & $0.93^{+0.13}_{-0.06}$ & 3.3 & -- & 2.8 & 1.7 & -- & 6.9 & 1.6 & -- & -- & 2.7 & -- & -- & -- \\
\rowcolor[gray]{0.9}
$\kappa_{b}$ & $0.89^{+0.14}_{-0.11}$ & 3.6 & 2.1 & 3.2 & 2.3 & 1.8 & 0.58 & 0.48 & 1.9 & 0.46 & 0.37 & 1.2 & 1.3 & 0.67 \\
$\kappa_{\mu}$ & $1.06^{+0.27}_{-0.30}$  & 4.6 & -- & 2.5 & 1.7 & 15 & 9.4 & 6.2 & 320$^\star$ & 13 & 5.8 & 8.9 & 10 & 8.9 \\
\rowcolor[gray]{0.9}
$\kappa_{\tau}$ & $0.92^{+0.13}_{-0.07}$ & 1.9 & 3.3 & 1.5 & 1.1 & 1.9 & 0.70 & 0.57 & 3.0 & 1.3 & 0.88 & 1.3 & 1.4 & 0.73 \\
\bottomrule
\end{tabular}
\end{adjustbox}
\caption{Current determinations and expected relative precision at $1 \sigma$ (\%) of the $\kappa$ parameters. Lack of sensitivity is indicated with a dash (-). The first column shows the current determinations obtained by ATLAS~\cite{ATLAS:2022vkf}. Note that in the fit all $\kappa$'s are assumed to be positive. The projections for HL-LHC (High-Luminosity LHC), LHeC (The Large Hadron Electron Collider), HE-LHC (The High-Energy Large Hadron Collider), ILC (The International Linear Collider), CLIC (The Compact Linear Collider), CEPC (The Circular Electron Positron Collider), and FCC-ee (The Future Circular Collider), are taken from Ref.~\cite{deBlas:2019rxi}.\label{Tab:Hcouplings}}
\end{table}
%
%Note that for the class of theories discussed in this paper $\tilde{\kappa}^f_i = 0$. 
The most stringent constraints on the Higgs couplings at $68\%$ confidence level, obtained by the ATLAS experiment~\cite{ATLAS:2022vkf}, are shown in Tab.~\ref{Tab:Hcouplings}. 

\begin{figure}
    \centering
    \includegraphics[width=0.95\linewidth]{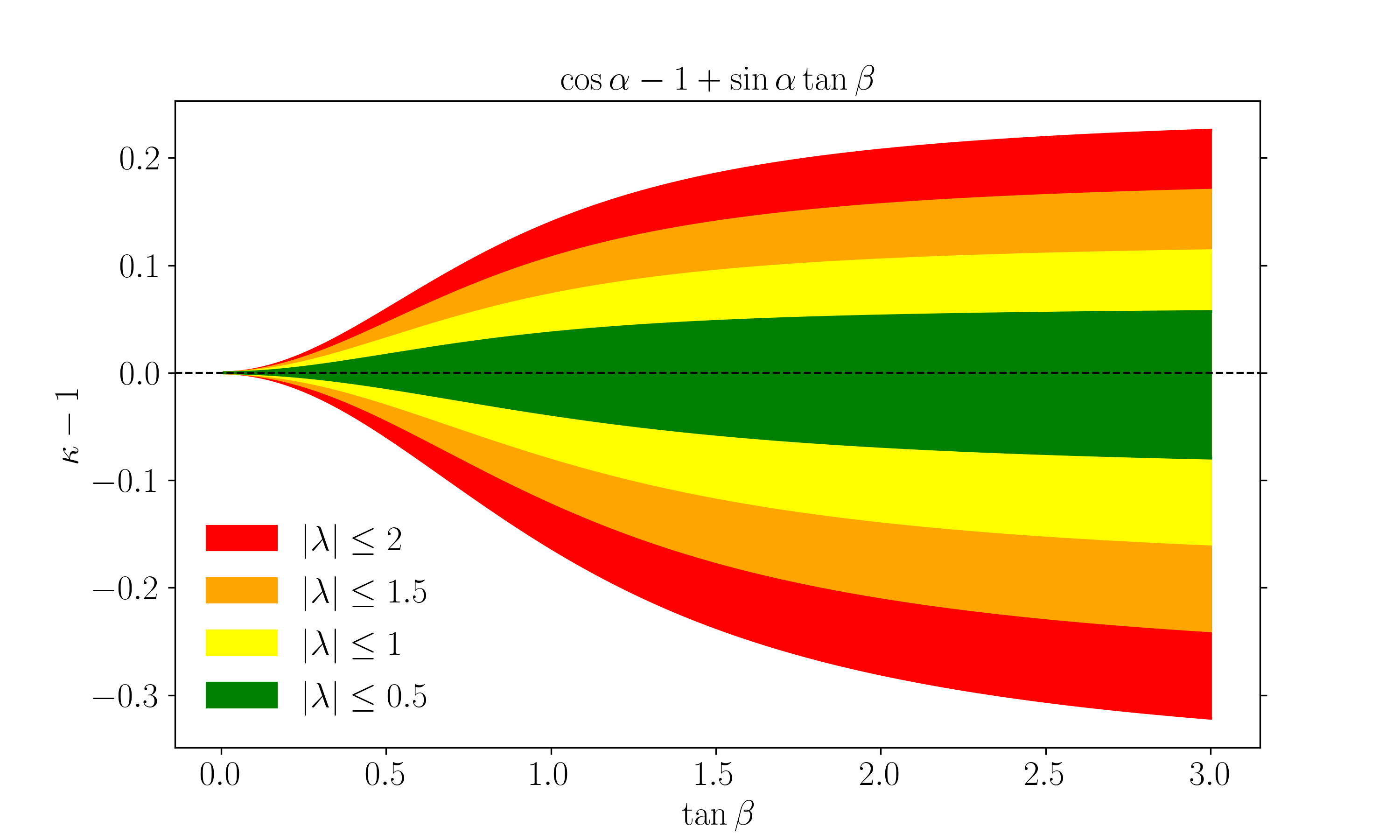}\\
    \includegraphics[width=0.95\linewidth]{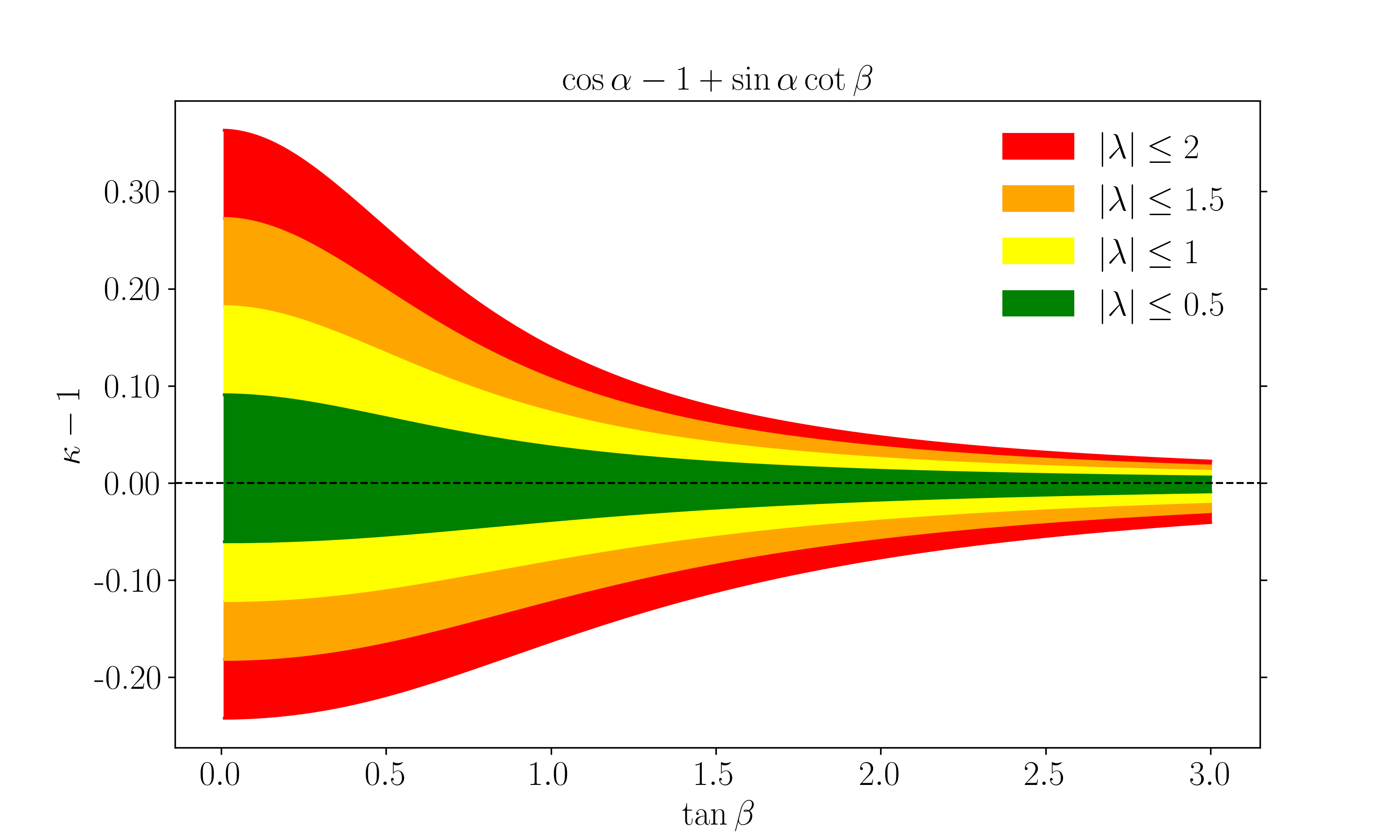}
    \caption{Illustrative expectations for the $\kappa$-factors in the 2HDM scenarios considered in this work, shown as a function of $\tan\beta$ and calculated using Eq.~\eqref{eq:kappa_minus_1} with $\hat{z}^{u,d.e}_{ii} /\hat{y}^{u,d,e}_{ii} = \tan \beta$ (upper panel) and $\hat{z}^{u,d.e}_{ii} /\hat{y}^{u,d,e}_{ii} = \cot \beta$ (lower panel). Each band corresponds to varying the quartic couplings, independently, within four different perturbative ranges, with $M = 1$ TeV held fixed.}
\label{fig:kappa_var_lambda}
\end{figure}

\begin{figure}
    \centering
    \includegraphics[width=0.95\linewidth]{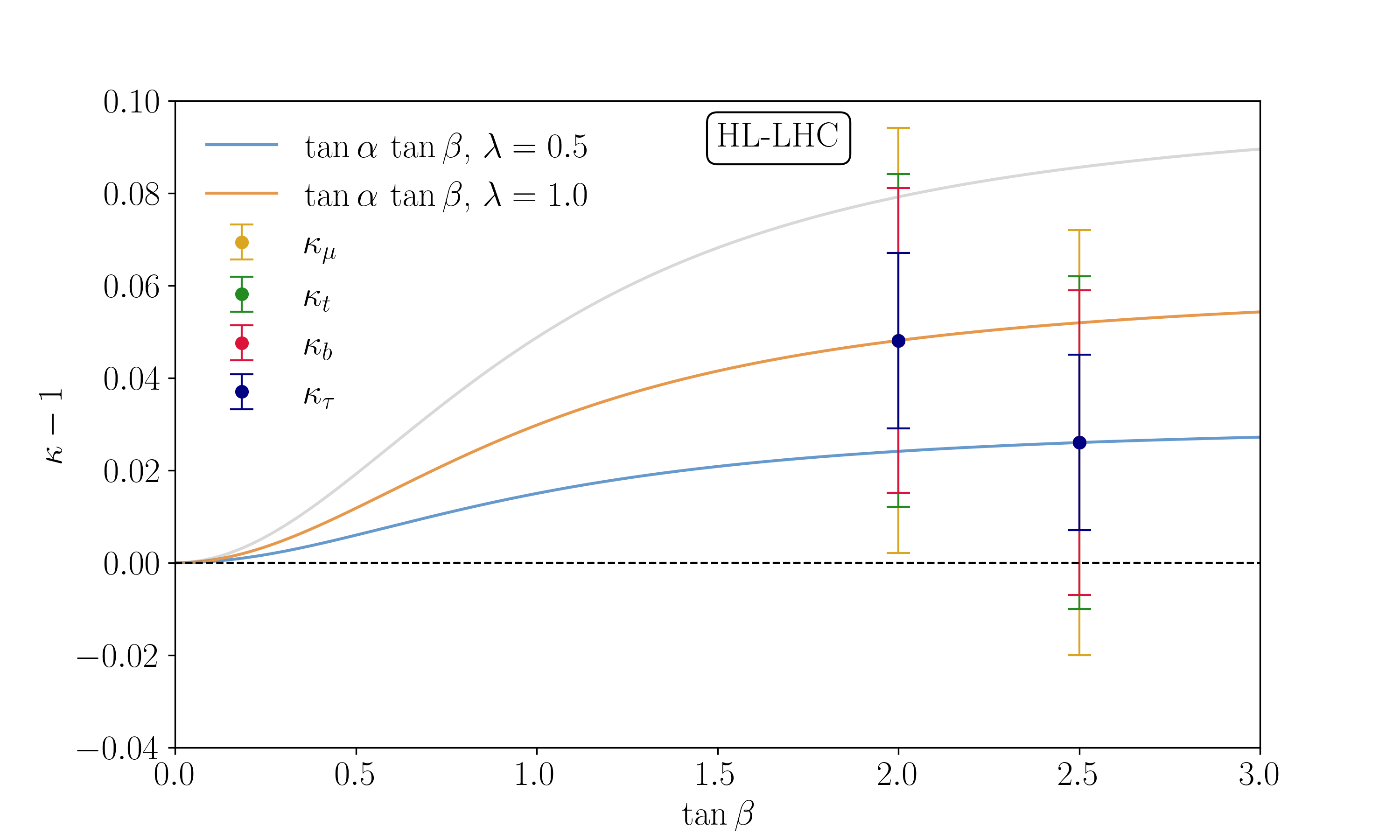}\\
    \includegraphics[width=0.95\linewidth]{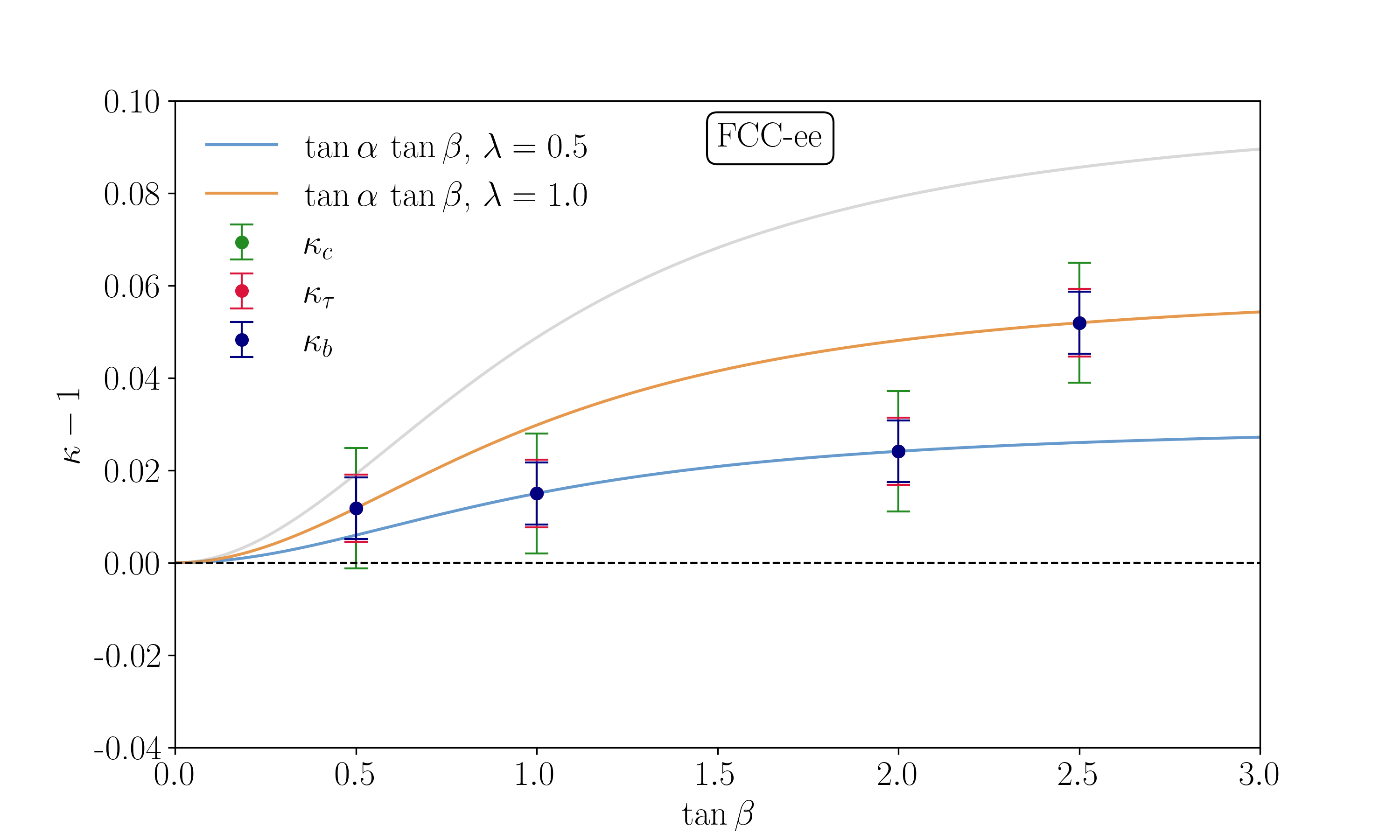}
    \caption{Illustrative expectations for the $\kappa$-factors in the 2HDM scenarios discussed in this work for $\hat{z}^{u,d.e}_{ii} /\hat{y}^{u,d,e}_{ii} = \tan \beta$, along with hypothetical future measurements at the HL-LHC and FCC-ee~\cite{deBlas:2019rxi}. Central values are chosen arbitrarily to highlight the potential discovery power of future sensitivities, shown at $68\%$ C.L.. Although shown as points, each represents a horizontal band. The parameter $\lambda$ appearing in the legends refer to the two combinations of quartics appearing in Eq.~\eqref{eq:mix_pert}, which are taken to be equal for illustrative purposes, $\lambda = \lambda_1 - \lambda_3 - \lambda_4 = -\lambda_2 + \lambda_3 + \lambda_4$. The region above the gray line is excluded by constraints on $D$ mixing for Model I. A similar exclusion line exists for Model II, but larger values of $\kappa -1$ may be allowed in other models.}
    \label{fig:kappa_future}
\end{figure}

Figure~\ref{fig:kappa_var_lambda} illustrates the predicted range for a representative $\kappa$ parameter as the quartic couplings are varied within the range, $[0.5, 2]$, with $M_H = 1\,\text{TeV}$ held fixed. The plot reveals that part of the parameter space is already being probed by current experiments. For clarity, the region excluded by $\Delta F = 2$ constraints is not shown, but it is implicitly understood. Although existing constraints are less stringent than those arising from $\Delta F = 2$ processes, upcoming experiments at the HL-LHC and future colliders are expected to significantly enhance sensitivity. In particular, they will enable precision tests of the Higgs couplings to bottom quarks, muons, and taus at the sub-percent level~\cite{deBlas:2019rxi,Black:2022cth}, as summarized in Table~\ref{Tab:Hcouplings}. Fig.~\ref{fig:kappa_future} shows the prediction for a representative $\kappa$ parameter as a function of $\tan\beta$ for two possible values of the combinations of quartics in Eq.~\eqref{eq:mix_pert}, taken to be equal for illustrative purposes, $\lambda = \lambda_1 - \lambda_3 - \lambda_4 = -\lambda_2 + \lambda_3 + \lambda_4$. Hypothetical measurements performed at HL-LHC and FCC-ee that showcase the potential discovery capabilities are also shown. Sensitivities are shown at $68\%$ C.L..

The off-diagonal couplings of $h$ to quarks are constrained from flavor-changing decays of the top quark~\cite{ATLAS:2018jqi, CMS:2021gfa}. The ATLAS searches for $t\to h c$ and $t\to h u$ give~\cite{ATLAS:2018jqi}
\begin{align}
\begin{split}
    \frac{\sin\alpha}{\sqrt{2}}\, \sqrt{\hat{z}^{u\, 2}_{23} + \hat{z}^{u\, 2}_{32}} \lesssim 0.055\,,\quad \frac{\sin\alpha}{\sqrt{2}}\, \sqrt{\hat{z}^{u\, 2}_{13} + \hat{z}^{u\, 2}_{31}} \lesssim 0.055\,.
\end{split}
\end{align}
Very similar results are obtained from CMS constraints~\cite{CMS:2021gfa}. As evident from Eq.~\eqref{eq:mix_pert}, these are not the leading constraints in this case either. Future colliders are expected to enhance the sensitivity of these measurements by a few orders of magnitude~\cite{Ake:2023xcz,Jueid:2024nbv, Zarnecki:2018wde, Ozsimsek:2022lte}, potentially rendering them competitive with the stringent constraints from $\Delta F = 2$ processes.

Furthermore, in these theories, the couplings of the SM Higgs to electroweak gauge bosons are modified with effects scaling as $(\lambda v^2 / M_H^2)^2$. Although future measurements at facilities such as FCC-ee and the muon collider are expected to reach per-mille-level precision, they are unlikely to provide the most stringent constraints.

\section{Conclusion}\label{sec:Conclusions}

We have established that a broad class of two-Higgs-doublet models, equipped with CP and Abelian flavor symmetries softly broken in the scalar potential, can naturally solve the strong CP problem. These theories are constructed so that $\bar{\theta}=0$ at tree-level and, remarkably, we discovered that the one-loop corrections necessarily vanish independent of the mass scale of the additional scalar doublet.
We explicitly verified the absence of one-loop contributions by analyzing the flavor structure of the radiative corrections to the Yukawa couplings in these theories. We further estimated two-loop corrections in specific models. While we cannot exclude the possibility of an exact cancellation of two-loop corrections, we find that, even in the absence of such cancellations, $\bar{\theta}$ remains below current experimental bounds, possibly leading to detectable signals in future measurements of the neutron EDM.

Importantly, our framework leads to concrete phenomenological implications. Flavor-changing neutral current contributions to $K, D$ and $B$ meson mixing give the strongest experimental bounds on the mass of the heavy Higgs doublet and the CP-even scalars mixing angle in the two examples we studied here. However, for {\it all} models in the framework, we have proved that the solution to the strong CP problem implies that {\it all} neutral scalar interactions conserve CP. Hence, there is essentially no bound from CP violation in meson mixing, and the constraints from meson mass splittings are typically near the TeV scale (and model-dependent). Hence, the additional scalar states, comprising a neutral CP-even, a CP-odd, and a charged Higgs boson, may give observable signals at current and future collider experiments.

Deviations from the SM in couplings of the light Higgs to $t, b, \tau, c$ and $\mu$ constitute a promising avenue for testing these models. Current limits from ATLAS and CMS experiments are already placing bounds on the size of the scalar quartic couplings, $\lambda_a$, and on $\tan \beta$ if the heavy Higgs mass is about 1 TeV. The High Luminosity run at the LHC is projected to increase the sensitivity to $t, b, \tau, \mu$ modes by a factor of 4-7. Barring accidental cancellations in the signal, this will allow a $2 \sigma$ signal for $\lambda_a \sim 1$, for most values of $\tan \beta$ consistent with the heavy Higgs being at 1 TeV. With FCC-ee projected to have 1\% sensitivities for $b, c, \tau$ modes, the $5 \sigma$ discovery reach for a heavy Higgs at 1 TeV is allowed for a large range of values of $\tan \beta$ and $\lambda_a \sim 1$.
Furthermore, the patterns of such deviations are highly constrained by the underlying flavor structure of the theory. Notably, these deviations exhibit characteristic correlations across different fermion channels, providing distinctive signatures that can be probed experimentally. 

The leading direct contributions to the neutron EDM, via 1-loop quark EDMs, are found to vanish in the two models explicitly studied, and the 2-loop contributions are found to be negligibly small. Furthermore, in all models the Weinberg operator is not generated at 1- or 2-loop level, due to the CP invariance of neutral scalar couplings, and can be safely neglected for all practical purposes. Nevertheless, in some models, for example our Model II, depending on values of unknown parameters such as $\tan \beta$, two-loop contributions to $\bar{\theta}$ can be as large as the present experimental bound. These theories will be significantly probed by the several experiments that aim to improve the sensitivity to the neutron EDM by two orders of magnitude over the next decade.  

Our framework is easily extended to the lepton sector by assigning $G_{HF}$ charges to the leptons.  Current and planned experiments to measure $\mu \to e \gamma$, $\mu \to e$ conversion and $\mu \rightarrow e e e$ aim to improve sensitivity to lepton flavor violation by up to 4 orders of magnitude. 
%Such processes are highly sensitive to new sources of flavor violation and 
This could provide new constraints or discovery prospects for 2HDM that solve the strong CP problem. 
However, predictions in the lepton sector are strongly model-dependent, especially as large neutrino mixing angles must be accommodated, and we leave a thorough investigation of the lepton sector and its phenomenological implications to future studies.

In conclusion, our results demonstrate that two-Higgs doublet models with softly broken CP and flavor symmetries are simple and remarkable candidates for addressing the strong CP problem. This motivates further theoretical investigations, particularly into the origins of soft CP and flavor breaking, as well as possible connections to other major puzzles in the scalar sector, such as the electroweak hierarchy problem. On the experimental front, as the extra Higgs doublet can have a mass near 1 TeV, these models exhibit a rich and distinctive phenomenology, motivating searches across multiple domains, from precision flavor and EDM measurements to Higgs physics. The prospect of discovering this scenarios in the coming decade underscores the importance of continued exploration on both theoretical and experimental grounds.

%%%%%%%%%%%%%%%%%%%%%%%%%%%%%%%%%%%%%%%
%%%%%%%%%%%%%%%%%%%%%%%%%%%%%%%%%%%%%%%
\newpage

\begin{center}
\textbf{\Huge Appendix}
\end{center}

\appendix
%\vspace{1.5cm}

 \section{Scalar Mass Matrix}\label{app:MassMatrix}

Here we provide the mass matrix for the two Higgs scalar potential of Eq.~\eqref{eq:scalarPot}. The relation between the flavor and Higgs basis, where only one scalar field $H_1$ has a non-zero vev $v^2=v_1^2+v_2^2$, is
\begin{align}
\begin{pmatrix}
H_1\\H_2
\end{pmatrix}
= \begin{pmatrix}
\cos\beta& - \sin\beta\, e^{-i\theta}\\
\sin\beta & \cos\beta\, e^{-i\theta}
\end{pmatrix} 
\begin{pmatrix}
\Phi_1\\
\Phi_2
\end{pmatrix}\,,
\end{align}
with $\cos{\beta} = v_1/v$ and $\sin{\beta} = -v_2/v$. We define
\begin{align}
\Phi_1 = \begin{pmatrix}
\Phi_1^+\\
\frac{v_1+\rho_1+i \eta_1}{\sqrt{2}}
\end{pmatrix}\,,\quad
\Phi_2 = e^{i\theta}\begin{pmatrix}
\Phi_2^+\\
\frac{v_2+\rho_2+i \eta_2}{\sqrt{2}}
\end{pmatrix}\,,\quad
H_1 = \begin{pmatrix}
G^+\\
\frac{v+H^0+i G^0}{\sqrt{2}}
\end{pmatrix}\,,\quad
H_2 = \begin{pmatrix}
H^+\\
\frac{R^0+i I^0}{\sqrt{2}}
\end{pmatrix}\,.
\end{align}
$G^{\pm}$ and $G^0$ are the Goldstone bosons associated with electroweak symmetry breaking. For the other scalars we find
\begin{align}
V \supset \mathcal{M}_{+}^2\, H^+H^- + 
\frac{1}{2}\begin{pmatrix}
H^0 & R^0 & I^0
\end{pmatrix}
\mathcal{M}
\begin{pmatrix}
H^0\\
R^0\\
I^0
\end{pmatrix} \, ,
\end{align}
with 
\begin{align}
\begin{split}
& \mathcal{M}_{+}^2 = \frac{\mu_{12}}{\sin\beta\cos\beta} - v^2\lambda_4\,,\\
& \mathcal{M}_{11}=2\,v^2(\cos^4\beta\, \lambda_1 + \sin^4\beta\, \lambda_2 + 2\cos^2\beta\sin^2\beta\,(\lambda_3 + \lambda_4))\,,\\
& \mathcal{M}_{12}=2\,v^2\,\cos\beta\sin\beta(\cos^2\beta (\lambda_1-\lambda_3-\lambda_4) + \sin^2\beta (-\lambda_2+\lambda_3+\lambda_4))\,,\\
& \mathcal{M}_{22} =2\, v^2 \cos^2\beta\sin^2\beta\, (\lambda_1 +\lambda_2 - 2 (\lambda_3 + \lambda_4)) + \frac{\mu_{12}}{\sin\beta \cos\beta} \,,\\
& \mathcal{M}_{13}=\mathcal{M}_{23}=0\,,\\
& \mathcal{M}_{33}= \frac{\mu_{12}}{\sin\beta\cos\beta}.
\end{split}
\label{eq:CPevenmassmatrix}
\end{align}

\section{Lagrangian in the \textit{Phase Basis}}\label{app:Lag}

We provide here the couplings of the scalars and charged weak gauge bosons to quarks in the unitary gauge and in the phase basis. %We start from Yukawa couplings of the form $- y^{d}\, H^{\dagger}_{\alpha} (q^{\alpha})^{T} \bar{d},\; - y^{u}\, H^{\alpha} \epsilon_{\alpha\beta} (q^{\alpha})^{T} \bar{u}$, where $\alpha$ is an SU(2) index, color and spinor indices are implicit and $q$ is transposed with respect to Weyl indices only. 
Using the conventions defined in Eqs.~\eqref{eq:Hbasis} and \eqref{eq:YukawaFlavorBasis}, as well as $q=(u,d)$ to denote left-handed quarks, we find : 
\begin{align}
\begin{split}
    % \mathcal{L} \; &\supset \;  -\frac{1}{\sqrt{2}}\left(u \, \bar{y}^u\, \bar{u} \,  + d \, \bar{y}^d\, \bar{d}  \right) H^0\, - \frac{1}{\sqrt{2}}\left(u \, \bar{z}^u\, \bar{u}  + d \, \bar{z}^d\, \bar{d} \, \right) R^0\\
    % &- \frac{i}{\sqrt{2}} \left(u \, \bar{z}^u\, \bar{u}  + d \, \bar{z}^d\, \bar{d} \right) A^0  + d \; V\,\bar{z}^u\, \bar{u} \; H^{+} - u \,  V^* \,\bar{z}^d\, \bar{d} \; H^{-} \\
    % & - \frac{g_2}{2} \left( u^\dagger\, V\, \bar{\sigma}_{\mu}\, d \; W^{+\, \mu} +  d^\dagger\,  \bar{\sigma}_{\mu}\, \, d\, Z^{\mu} + u^\dagger \,  \bar{\sigma}_{\mu}\, u \, Z^{\mu}\right) + {\rm h.c.}
    \mathcal{L_Y} \; \supset & \;  \frac{1}{\sqrt{2}}\left(u \, \bar{y}^u\, \bar{u} \,  + d \, \bar{y}^d\, \bar{d}  \right) H^0\, + \frac{1}{\sqrt{2}}\left(u \, \bar{z}^u\, \bar{u}  + d \, \bar{z}^d\, \bar{d} \, \right) R^0+ \frac{i}{\sqrt{2}} \left(u \, \bar{z}^u\, \bar{u}  + d \, \bar{z}^d\, \bar{d} \right) I^0\\
    &  + d \; V\,\bar{z}^u\, \bar{u} \; H^{+}- \frac{g_2}{2} %\left( 
    u^\dagger\, V\, \bar{\sigma}_{\mu}\, d \; W^{+\, \mu} %+  d^\dagger\,  \bar{\sigma}_{\mu}\, \, d\, Z^{\mu} + u^\dagger \,  \bar{\sigma}_{\mu}\, u \, Z^{\mu}\right) 
    + {\rm h.c.}
\end{split}
\end{align}
Real matrices are indicated by an overbar 
\begin{align}
\begin{split}
    \bar{y}^{u,d} = P^{u,d}\, y^{u,d} \, P^{\bar{u},\bar{d}}\ ,\\
    \bar{z}^{u,d} = P^{u,d}\, z^{u,d} \, P^{\bar{u},\bar{d}}\ ,
\end{split}
\end{align}
where $P^{u,d,\bar{u},\bar{d}}$ are diagonal phase matrices and $y^{u,d},z^{u,d}$ are Higgs basis matrices. $V= {P^u}^*P^d$ is a diagonal matrix of phases that is the phase-basis-equivalent of the CKM matrix. Coupling matrices for the neutral gauge bosons are real and proportional to the unit matrix.

%\section{CP-even Higgs interactions}\label{app:CPNC}
\section{Yukawa Interactions of All Neutral Scalars Conserve CP}\label{app:CPNC}

From a direct analysis of many 2HDM where softly-broken CP and $\mathbb{Z}_n$ flavor symmetries solve the strong CP problem, we found that all neutral Higgs couplings to quarks conserve CP. In this appendix, we give a general model-independent proof of this result. More precisely, it is possible to show the equivalence between two sets of 2HDM :
\begin{enumerate}
%\item \label{item2} Vanishing entries are always linked to the flavor symmetry, not to tuning, when we scan over models.
\item\label{item3}\textit{Models that solve the strong CP via flavor.} The determinant of a given Yukawa matrix combined with the Higgs fields, say $x^\alpha\Phi_\alpha$, is proportional to a single monomial $\Phi_1^{n_1}\Phi_2^{n_2}$, with $n_1 + n_2 =3$, for arbitrary values of the allowed $x^\alpha_{ij}$ couplings. %No tuning is allowed when building models with that property. 
A natural cancellation between $\arg \det m_u$ and $\arg \det m_d$ then solves the strong CP problem.
\item\label{item4} \textit{Models with CP conservation in the neutral scalar sector.} Treating $u$ and $d$ independently, i.e. not as a single doublet $q=(u,d)$, the quark fields can be rephased to make the Yukawa couplings of all neutral scalars real and therefore CP-conserving.\footnote{Indeed, starting from these real matrices, the quark mass basis can be reached through an orthogonal transformation of the quark fields. The only complex components are to be found in terms which couple up and down quarks to the W boson and charged Higgses. In order to deduce from this that there is no CP violation arising from neutral Higgs exchange, it is also important that, in our models, there is no CP violation in the Higgs potential in the Higgs basis.}
\end{enumerate}

\noindent Since these properties hold for the up and down sectors independently, we focus on the up sector and the Yukawa matrix $x$, but the exact same arguments hold for the down sector and $\tilde x$. For this paper, the most interesting aspect is the direct implication (1.$\implies$2.), so we do not report the reciprocal proof and leave it as an exercise for the reader. 

Reasoning by contraposition, let us assume that it is not possible to rephase $u$ and $\bar{u}$ so as to make $x^\alpha_{ij}U_{\alpha\beta}$ real for $\beta=1,2$ at once. From the expression of $U$ in Eq.~\eqref{eq:Umatrix}, we see that $U_{\alpha 1}$ and $U_{\alpha 2}$ have the same phase and, for a given pair $(i,j)$, $x_{ij}^1$ and $x_{ij}^2$ cannot be both non-zero, since we choose different flavor charges for the two Higgses. Thus, $x^\alpha_{ij}U_{\alpha\beta}$ has the phase of the only $U_{\alpha 1}$ for which $x^\alpha_{ij}$ does not vanish, if any. Therefore, we cannot make $x^\alpha_{ij}U_{\alpha\beta}$ real through rephasings if and only if we cannot make $x^\alpha_{ij}U_{\alpha 1}$ real. Defining $M_{ij}=x^\alpha_{ij}U_{\alpha 1}$ (so $M=\frac{\sqrt 2}{v}m_u$), one can measure the impossibility to redefine all phases away from $M$ via appropriate rephasing invariants. In our case, those are $\text{Im}(M_{ij}M_{il}^*M_{kl}M_{kj}^*)$ and $\text{Im}(M_{ij}M_{il}^*M_{kl}M_{km}^*M_{nm}M_{nj}^*),$\footnote{\label{footnote:invariants}Let us prove that these are the only two independent rephasing-invariant combinations of $M_{ij}$ elements, by going through their explicit construction. Without loss of generality, the first one can be chosen to be $M_{11}$. To absorb the phase shifts of the first index, there must appear in the product some $M_{1j}^*$, where again we can choose $j=2$. Then, some $M_{i2}$ needs to appear, where we can choose $i=2$. ($i=1$ would make $|M_{12}|^2$ appear among the terms we already mentioned, which could be taken out of the imaginary part.) We arrive at an object $M_{11}M_{12}^*M_{22}$ which rephases like $M_{21}$. We can either make this invariant by multiplying it with $M_{21}^*$, which yields the first kind of invariants that we considered, or we can cancel the first index phase shift with $M_{23}^*$. (Again, using $M_{22}^*$ would allow us to extract a real quantity.) Taking care of the second index can be done either by multiplication of $M_{13}$ or $M_{33}$. (Not $M_{23}$ for yet the same reason.) The former yields a product which transforms as the initial $M_{11}$, meaning that we are in the process of building an object which will factorize in several rephasing invariants. With the latter, we form $M_{11}M_{12}^*M_{22}M_{23}^*M_{33}$ which transforms as $M_{31}$. Closing with $M_{31}^*$ yields our second invariant, otherwise the only other option is to use $M_{32}^*$ (not $M_{33}^*$), in which case we end up with an object which transforms like $M_{11}M_{12}^*$, which we already encountered. This ends the proof.}  one of which should be non-zero by assumption.

Let us start by assuming that it is the latter, since, as we will show below, it yields the only relevant case for a $\mathbb{Z}_n$ flavor symmetry with odd $n$, such as the $\mathbb{Z}_3$ considered in Sec.~\ref{sec:Models}. Thanks to the freedom of redefining the lines and columns of $M$ (i.e. of reordering $u$ and $\bar{u}$), which only affects the determinant by a sign (and therefore does not affect the number of monomials constituting it), we can always choose the invariant to be $\text{Im}(M_{11}M_{12}^*M_{22}M_{23}^*M_{33}M_{31}^*)={\cal I}$ (see foonote~\ref{footnote:invariants}). Since $U_{\alpha 1}=\frac{\sqrt{2}\langle \Phi_\alpha\rangle}{v}$, we can define $N_{ij}=x^\alpha_{ij} \Phi_\alpha$ (so that $M=(v/\sqrt{2})^{-1}\langle N \rangle$) and think about $\cal I$ computed with $M$ replaced by $N$. We then see that $\cal I$, to be non-zero, must at least receive one contribution proportional to $\Phi_1$. We can choose it to be $N_{11}$, up to redefinitions of rows and columns, and potentially a conjugation, of $N$. Actually, at least two $\Phi_1$ are required by the flavor symmetry. Indeed, denoting $q_\psi$ the charge of a field $\psi$, the non-vanishing nature of the various entries forming ${\cal I}$ would imply that
\begin{align}
\begin{split}
    0&=q_{q_1}+q_{\bar u_1}+q_{q_2}+q_{\bar u_2}+q_{q_3}+q_{\bar u_3}-(q_{q_1}+q_{\bar u_2}+q_{q_2}+q_{\bar u_3}+q_{q_3}+q_{\bar u_1})\\
    &= q_{\Phi_2}-q_{\Phi_1} \text{ mod } n \ \text{ if $N$ had texture } \left(\begin{matrix} \Phi_1 & \Phi_2 & * \\ * & \Phi_2 & \Phi_2 \\ \Phi_2 & * & \Phi_2 \end{matrix}\right)\ ,\\
\end{split}
\end{align}
where $*$ could be either $0$ or one of the Higgs fields. For the flavor symmetry to distinguish the two Higgses, we want that $q_{\Phi_2}\neq q_{\Phi_1} \text{ mod } n$, thus such a texture is not allowed. If two $\Phi_1$ contribute to $\cal I$, its non-zero nature imposes that the second $\Phi_1$ does not appear in $N_{12,23,31}$, so that we can choose it to be found in $N_{22}$, up to a cyclic shift of all rows and columns. Finally, if there are three $\Phi_1$, distinguishing the Higgses through the flavor symmetry imposes that they sit in $N_{11,22,33}$. Four, five and six $\Phi_1$ are treated identically, since they correspond to two, one and zero $\Phi_2$. We are thus left with three structures,
 \begin{equation}
    \left(\begin{matrix} \Phi_1 & \Phi_1 & * \\ * & \Phi_1 & \Phi_2 \\ \Phi_2 & * & \Phi_1 \end{matrix}\right) \ , \quad \left(\begin{matrix} \Phi_1 & \Phi_2 & * \\ * & \Phi_1 & \Phi_2 \\ \Phi_2 & * & \Phi_1 \end{matrix}\right) \ , \quad \left(\begin{matrix} \Phi_1 & \Phi_2 & * \\ * & \Phi_1 & \Phi_2 \\ \Phi_2 & * & \Phi_2 \end{matrix}\right) \ .
\end{equation}
Replacing the arbitrary coefficients by $0$, which is allowed as we bar tuning between different contributions to the determinant, we see by direct computation that the three determinants contain more than one monomial.

Finally, we consider the remaining case where the non-vanishing rephasing invariant is $\text{Im}(N_{11}N_{12}^*N_{21}^*N_{22})={\cal I}$. We then see that the $2\times 2$ block 
\begin{equation}
    \left(\begin{matrix} N_{11} & N_{12} \\ N_{21} & N_{22}\end{matrix}\right)
\end{equation}
has a determinant which must be formed out of two different monomials. Indeed, only the following three assignements (up to reshuffling of the first two lines and columns) yield a non-zero ${\cal I}$, 
\begin{equation}
\label{eq:textureNoRephasing1}
    \left(\begin{matrix} \Phi_1 & \Phi_2 \\ \Phi_2 & \Phi_2 \end{matrix}\right) \, , \quad \left(\begin{matrix} \Phi_1 & \Phi_2 \\ \Phi_1 & \Phi_1 \end{matrix}\right) \ , \quad \left(\begin{matrix} \Phi_1 & \Phi_2 \\ \Phi_2 & \Phi_1 \end{matrix}\right)  \ .
\end{equation}
Actually, for the flavor symmetry to distinguish the two Higgses, only the third texture %in Eq.~\eqref{eq:textureNoRephasing1} 
is compatible with a $\mathbb{Z}_n$ flavor symmetry, and only with $n$ even :
\begin{equation}
    0=q_{q_1}+q_{\bar u_1}+q_{q_2}+q_{\bar u_2}-(q_{q_1}+q_{\bar u_2}+q_{q_2}+q_{\bar u_1})= 2(q_{\Phi_2}-q_{\Phi_1}) \text{ mod } n\ \text{ for }\left(\begin{matrix} \Phi_1 & \Phi_2 \\ \Phi_2 & \Phi_1 \end{matrix}\right)\, .
\end{equation}
For a flavor symmetry with $n$ odd, there is no such quartic rephasing invariant. Due to hypercharge invariance, we can fix $q_{\Phi_1}=0$, which we do in the following, then $q_{\Phi_2}=n/2$. 

$N$ being a full rank matrix (since its vev is proportional to the up-type Yukawa matrix), we must either have $N_{33}\neq 0$, or $N_{3i}\neq 0$ and $N_{j3}\neq 0$ for $i,j=1$ or $2$. In the former case, $\det N$ contains at least two different monomials and the proof ends. In the latter case, it also turns out that $N_{33}\neq 0$. Let us assume that $i=1$ without loss of generality, as the third texture in Eq.~\eqref{eq:textureNoRephasing1} is invariant under the exchange of first and second rows and columns. Let us also denote $\Phi_{ab}$ the Higgs field which appears in $N_{ab}$. Then,
\begin{equation}
q_{q_3}+q_{\bar u_1}+q_{\Phi_{31}}=0\text{ mod } n=q_{q_j}+q_{\bar u_3}+q_{\Phi_{j3}} \,.
\end{equation}
Thus,
\begin{equation}
\label{eq:N33charge1}
q_{q_3}+q_{\bar u_3}=q_{q_3}+q_{\bar u_1}-\left(q_{q_j}+q_{\bar u_1}\right)+q_{q_j}+q_{\bar u_3}=-q_{\Phi_{31}}+q_{\Phi_{j1}}-q_{\Phi_{j3}} \text{ mod } n \ ,
\end{equation}
and, for any $(k,l)$ such that $N_{k1},N_{jl},N_{kl}\neq 0$,
\begin{align}
\begin{split}
q_{q_3}+q_{\bar u_3}&=q_{q_3}+q_{\bar u_1}-\left(q_{q_k}+q_{\bar u_1}+q_{q_j}+q_{\bar u_l}-\left[q_{q_k}+q_{\bar u_l}\right]\right)+q_{q_j}+q_{\bar u_3}\\
&=-q_{\Phi_{31}}+q_{\Phi_{k1}}+q_{\Phi_{jl}}-q_{\Phi_{kl}}-q_{\Phi_{j3}} \text{ mod } n \,.
\end{split}
\label{eq:N33charge2}
\end{align}
If $j=1$, we can choose $k=l=2$ and find from Eq.~\eqref{eq:N33charge2} that  
\begin{equation}
q_{q_3}+q_{\bar u_3}=2q_{\Phi_2}-q_{\Phi_{31}}-q_{\Phi_{j3}} \text{ mod } n=(2q_{\Phi_2}\,,\,q_{\Phi_2} \,\text{or } 0) \text{ mod } n=(0 \,\text{ or } -q_{\Phi_2}) \text{ mod } n\,,
\end{equation} 
depending on the four possible assignments of $\Phi_{31},\Phi_{j3}$. At the same time, from Eq.~\eqref{eq:N33charge1},
\begin{equation}
q_{q_3}+q_{\bar u_3}=-q_{\Phi_{31}}-q_{\Phi_{j3}} \text{ mod } n=(0\,,\,-q_{\Phi_2} \,\text{or } -2q_{\Phi_2}) \text{ mod } n=(0 \,\text{ or } -q_{\Phi_2}) \text{ mod } n \ ,
\end{equation}
hence $N_{33}\neq 0$ since $q_3\bar u_3$ can couple to $\Phi_1$ or $\Phi_2$. Similarly, if $j=2$, we can choose $k=1,l=2$ to obtain the same result.
% \begin{equation}
% q_{q_3}+q_{\bar u_3}=-q_{\Phi_2}-q_{\Phi_{31}}-q_{\Phi_{j3}} \text{ mod } n=(-q_{\Phi_2}\,,\, -2q_{\Phi_2}\,\text{or } -3q_{\Phi_2}) \text{ mod } n\,,
% \end{equation} 
% and
% \begin{equation}
% q_{q_3}+q_{\bar u_3}=q_{\Phi_2}-q_{\Phi_{31}}-q_{\Phi_{j3}} \text{ mod } n=(q_{\Phi_2}\,,\, 0\,, \,\text{or }  -q_{\Phi_2}) \text{ mod } n \ ,
% \end{equation}
% so that  and $q_{q_3}+q_{\bar u_3}=0$ or $-q_{\Phi_2}$, and again $N_{33}\neq 0$. 
This concludes the proof.

\section{Two-loop Contributions to $\bar{\theta}$}\label{app:2loops}

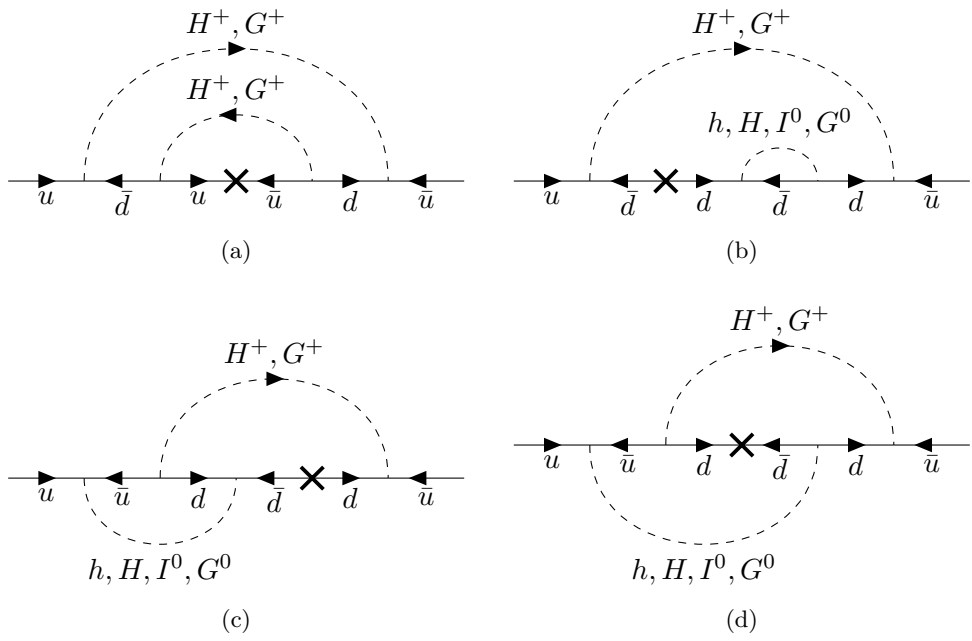
\begin{figure}
\centering    
\subfloat[]{
\begin{tikzpicture}[scale=0.5]
    \begin{feynman}
        \vertex (L3) at (-6,0);
        \vertex (L2) at (-4,0);
        \vertex (L1) at (-2,0);
        \vertex (R1) at (+2,0);
        \vertex (R2) at (+4,0);
        \vertex (R3) at (+6,0);
        \vertex (M) at (0,0);
        \vertex (T2) at (0,4);
        \vertex (T1) at (0,2);
        %\vertex (B) at (0,-2) {$\langle H_1 \rangle$};
        \node[cross out, draw, minimum size=7pt, inner sep=0pt, line width=0.5mm] at (0, 0) {};
        \diagram*{
            (L3) -- [fermion,edge label' = {$u$}] (L2),
            (L2) -- [anti fermion, edge label' = {$\bar d$}] (L1),
            (L1) -- [ fermion, edge label' = {$u$}] (M),
            (M) -- [anti fermion, edge label' = {$\bar u$}] (R1),
            (R1) -- [ fermion, edge label' = {$ d$}] (R2),
            (R2) -- [anti fermion, edge label' = {$\bar u$}] (R3),
            (L2) -- [charged scalar, half left, edge label={$H^+, G^+$}] (R2),
            (L1) -- [anti charged scalar, half left, edge label={$H^+, G^+$}] (R1)
            };
    \end{feynman}
    \end{tikzpicture}
}\hspace{0.1in}
\subfloat[]{
\begin{tikzpicture}[scale=0.5]
    \begin{feynman}
        \vertex (L3) at (-6,0);
        \vertex (L2) at (-4,0);
        \vertex (L1) at (-2,0);
        \vertex (R1) at (+2,0);
        \vertex (R2) at (+4,0);
        \vertex (R3) at (+6,0);
        \vertex (M) at (0,0);
        \vertex (T2) at (0,4);
        \vertex (T1) at (0,2);
        %\vertex (B) at (0,-2) {$\langle H_1 \rangle$};
        \node[cross out, draw, minimum size=7pt, inner sep=0pt, line width=0.5mm] at (-2, 0) {};
        \diagram*{
            (L3) -- [fermion,edge label' = {$u$}] (L2),
            (L2) -- [anti fermion, edge label' = {$\bar d$}] (L1),
            (L1) -- [ fermion, edge label' = {$d$}] (M),
            (M) -- [anti fermion, edge label' = {$\bar d$}] (R1),
            (R1) -- [ fermion, edge label' = {$d$}] (R2),
            (R2) -- [anti fermion, edge label' = {$\bar u$}] (R3),
            (M) -- [ scalar, half left, edge label={$h, H, I^0, G^0$}] (R1),
            (L2) -- [charged scalar, half left, edge label={$H^+, G^+$}] (R2)
            };
    \end{feynman}
\end{tikzpicture}
}

\subfloat[]{
\begin{tikzpicture}[scale=0.5]
    \begin{feynman}
        \vertex (L3) at (-6,0);
        \vertex (L2) at (-4,0);
        \vertex (L1) at (-2,0);
        \vertex (R1) at (+2,0);
        \vertex (R2) at (+4,0);
        \vertex (R3) at (+6,0);
        \vertex (M) at (0,0);
        \vertex (T2) at (0,4);
        \vertex (T1) at (0,2);
        %\vertex (B) at (0,-2) {$\langle H_1 \rangle$};
        \node[cross out, draw, minimum size=7pt, inner sep=0pt, line width=0.5mm] at (+2, 0) {};
        \diagram*{
            (L3) -- [fermion,edge label' = {$u$}] (L2),
            (L2) -- [anti fermion, edge label' = {$\bar u$}] (L1),
            (L1) -- [ fermion, edge label' = {$d$}] (M),
            (M) -- [anti fermion, edge label' = {$\bar d$}] (R1),
            (R1) -- [ fermion, edge label' = {$d$}] (R2),
            (R2) -- [anti fermion, edge label' = {$\bar u$}] (R3),
            (L2) -- [ scalar, half right, edge label'={$h, H, I^0, G^0$}] (M),
            (L1) -- [charged scalar, half left, edge label={$H^+, G^+$}] (R2)
            };
    \end{feynman}
\end{tikzpicture}
}\hspace{0.1in}
\subfloat[]{
\begin{tikzpicture}[scale=0.5]
    \begin{feynman}
        \vertex (L3) at (-6,0);
        \vertex (L2) at (-4,0);
        \vertex (L1) at (-2,0);
        \vertex (R1) at (+2,0);
        \vertex (R2) at (+4,0);
        \vertex (R3) at (+6,0);
        \vertex (M) at (0,0);
        \vertex (T2) at (0,4);
        \vertex (T1) at (0,2);
        %\vertex (B) at (0,-2) {$\langle H_1 \rangle$};
        \node[cross out, draw, minimum size=7pt, inner sep=0pt, line width=0.5mm] at (0, 0) {};
        \diagram*{
            (L3) -- [fermion,edge label' = {$u$}] (L2),
            (L2) -- [anti fermion, edge label' = {$\bar u$}] (L1),
            (L1) -- [  fermion, edge label' = {$d$}] (M),
            (M) -- [anti fermion, edge label' = {$\bar d$}] (R1),
            (R1) -- [fermion, edge label' = {$d$}] (R2),
            (R2) -- [anti fermion, edge label' = {$\bar u$}] (R3),
            (L2) -- [ scalar, half right, edge label'={$h, H, I^0, G^0$}] (R1),
            (L1) -- [charged scalar, half left, edge label={$H^+, G^+$}] (R2)
            };
    \end{feynman}
\end{tikzpicture}
}

\caption{Examples of two-loop radiative corrections to the up-type quark mass matrix due to the exchange of charged Higgs bosons, or one charged and one neutral. Diagrams with similar topologies but only neutral scalars do not violate $CP$ or contribute to $\bar\theta$. 
% Diagrams with a cubic scalar vertex only exist among neutral Higgses which do not mediate CP violation, while quartic scalar vertices contribute at three loops.
}
\label{fig:twoloopFD}
\end{figure}

The two-loop corrections to the quark Yukawa couplings that could reintroduce $\bar{\theta}$ are shown in Fig.~\ref{fig:twoloopFD}. The general form of the amplitude of each diagram is
\begin{align}
    \mathcal{M} = \frac{\mathcal{A}}{(16\pi^2)^2}\bigg(1+\frac{1}{\epsilon}+f(m_1,m_2)\bigg)\,,
\end{align}
where $\mathcal{A}$ is the flavor part of the diagram and involves the product of five Yukawa matrices and $f(m_1,m_2)$ is a dimensionless function of the two scalar masses in the loops. Using the unitarity of the $U$ matrix, we prove that the $f$-independent part of each diagram does not change the phases of the determinants of $y^u$ and $y^d$. 

We first show the explicit calculation for diagram (a). For the mass-independent piece we find
\begin{align}
\begin{split}
    \frac{\mathcal{A}}{(16\pi^2)^2}\bigg(1+\frac{1}{\epsilon} \bigg) &= \frac{1}{(16\pi^2)^2}\bigg(1+\frac{1}{\epsilon} \bigg)\sum_{i,j} \tilde{x}^\alpha \tilde{x}^{\beta \dagger} x^\gamma x^{\delta \dagger} x^\varepsilon
U^*_{\alpha i} U_{\beta j} U_{\gamma_1} U^*_{\delta j} U_{\varepsilon i}\\
%&= \frac{1}{(16\pi^2)^2}\bigg(1+\frac{1}{\epsilon} \bigg)\tilde{x}^\alpha \tilde{x}^{\beta \dagger} x^\gamma x^{\delta \dagger} x^\varepsilon\delta_{\alpha \varepsilon} \delta_{\beta \delta} U_{\gamma_1}\\
&= \frac{1}{(16\pi^2)^2}\bigg(1+\frac{1}{\epsilon} \bigg)\tilde{x}^\alpha \tilde{x}^{\beta \dagger} x^\gamma x^{\beta} x^{\alpha} U_{\gamma_1}\,.
\end{split}
\end{align}
The same argument that we used for one-loop diagrams applies: the last term has the same structure as $y^u$ since $\tilde{x}^\alpha \tilde{x}^{\beta \dagger} x^\gamma x^{\beta} x^{\alpha}$ has the same flavor charges as $x^\beta$ and therefore $\arg\det (y^u+\delta y^u_{\rm 2loops}) = \arg\det (y^u)$. It is easy to show that the same result applies to each diagram. For the $f$-dependent part of the amplitude, one would need to compute the two-loop integrals. Instead, we follow a different approach. In the main text, we present two illustrative theories and we estimate each contribution to $\bar{\theta}$ assuming $f(m_i,m_j)\sim 1$, finding that they are all naturally below the experimental limit. While a dedicated scan of two-loop contributions in all $\mathbb{Z}_3$ theories is beyond the scope of this work, we anticipate that they will generically be small enough to pass experimental constraints.

%\newpage
%\bibliographystyle{ieeetr}
\bibliographystyle{JHEP}
\bibliography{biblio.bib}

\end{document}